\documentclass{vldb}

\usepackage{pdfpages}
\usepackage[linesnumbered,ruled,vlined]{algorithm2e}
\usepackage{algpseudocode}
\usepackage{amsmath}
\usepackage{amssymb}

\usepackage{wrapfig,epsfig}
\usepackage{balance}
\usepackage{listings}
\usepackage{epstopdf}
\usepackage{subfigure}
\usepackage{array}
\usepackage{times}
\usepackage{lipsum}
\usepackage{url}
\usepackage{soul}

\usepackage{xcolor}
\usepackage{framed}
\usepackage{lipsum}

\colorlet{shadecolor}{blue!20}

\DeclareMathOperator*{\argmin}{arg\,min}

\newtheorem{example}{Example}
\newtheorem{lemma}{Lemma}
\newtheorem{property}{Property}
\newtheorem{definition}{Definition}

\newtheorem{theorem}{Theorem}

\def\done{\hspace*{\fill} $\framebox[2mm]{}$}

\begin{document}

\title{Finding Multiple New Optimal Locations in a Road Network}

\numberofauthors{3}

\author{
\alignauthor
Ruifeng Liu\\
\affaddr{The Chinese University of Hong Kong}\\
\email{rfliu@cse.cuhk.edu.hk}
\alignauthor
Ada Wai-Chee Fu\\
\affaddr{The Chinese University of Hong Kong}\\
\email{adafu@cse.cuhk.edu.hk}
\alignauthor
Zitong Chen\\
\affaddr{The Chinese University of Hong Kong}\\
\email{ztchen@cse.cuhk.edu.hk}
\and
\alignauthor
Silu Huang\\
\affaddr{Univeristy of Illinois Urbana-Champagne}\\
\email{shuang86@illinois.edu}
\alignauthor
Yubao Liu\\
\affaddr{Sun Yat Sen University}\\
\email{liuyubao@mail.sysu.edu.cn}
}

\maketitle

\begin{abstract}
We study the problem of optimal location querying for location-based services in road networks, which aims to find locations for new servers or facilities.
The existing optimal solutions on this problem consider only the cases with one new server. When two or more new servers are to be set up, the problem with minmax cost criteria, MinMax, becomes NP-hard. 
In this work we identify some useful properties about the potential locations for the new servers, from which we
derive a novel algorithm for MinMax, and show that
it is efficient when the number of new servers is small.
When the number of new servers is large, we propose an efficient 3-approximate algorithm.
We verify with experiments on real road networks that our solutions are effective and attains significantly better result quality compared to the existing greedy algorithms. 
\end{abstract}

\section{INTRODUCTION}
\label{sec:intro}

We study the problem of optimal location querying (OLQ) in a road network. In this problem we are given a set $C$ of weighted clients and a set $S$ of servers in a road network $G = (V, E)$, where $V$ is a set of vertices and $E$ is a set of edges. We are also given a set $E^0$ of eligible edges in the graph where new servers can be built. The objective of this query is to identify a set of $k$ ($k \geq 1$) locations (points) on $E^0$ such that $k$ new servers set up at these locations can optimize a certain cost function, which is defined based on the distances between the servers and the clients. Here, we focus on the minmax cost function.
The problem is how to add new servers on top of the existing servers so that the maximum weighted distance from any client to its nearest server is minimized. We call such a problem
a MinMax problem, or simply MinMax.

Solutions to the MinMax location problem have been found to be useful in various applications such as location planning and location-based marketing, and have been studied in recent works \cite{xiao2011optimal,chen2014efficient,chen2015OLQrep}.
Consider the case of a pizza chain that plans to build a new outlet in the city, aiming to minimize the worst case delivery time to clients. Given the set $S$ of existing pizza outlets and the set $C$ of all clients, the company wants to find a location for the new outlet on a set of eligible road segments so that the maximum distance from any client to its nearest outlet is minimized.
This is an example of MinMax. Another example is in city planning for new hospitals; we may set an objective to minimize the maximum distance from any residence to its nearest hospital.
Solutions to MinMax are useful for the location selection for facilities in general, such as schools, libraries, sports facilities, fire stations, post offices, police stations, etc.

Xiao el al. \cite{xiao2011optimal} first investigated the MinMax location query problem in road networks.
They considered the problem with a single new server.
Their solution adopts a divide-and-conquer paradigm, and has a
time complexity of $O(|V|+|S|+|C|)^{2}log(|V|+|S|+|C|)$.
%
%
Recently, an improved algorithm for Minmax location query was proposed in \cite{chen2014efficient}.
Based on a new concept called the nearest location component, their proposed algorithm can typically reduce much of the search space.
The time complexity of their algorithm is $O(m|V|log|V|+|V||C|log|C|)$, where $m$ is typically much smaller than $|C|$. The run time is significantly less than that of the algorithm for Minmax location query in \cite{xiao2011optimal}.

However, the above solutions assume that only one new server is to be added, and become inapplicable when more servers are to be built.
For real-life applications, such as a pizza delivery chain store,
there can be planning of multiple new outlets in a city for a
fiscal year. We may also have requirements of multiple new locations for public facilities.
While the minmax problem for a single new server is tractable, it becomes
intractable for two or more new servers.
The multi-server version of the MinMax problem is shown to be NP-hard in \cite{chen2015OLQrep}. A greedy algorithm is proposed which applies the single server solution repeatedly until $k$ servers are settled.
However, results from the greedy algorithm can be arbitrarily poor. Our
empirical studies show that the greedy approach has poor approximation ratio when compared to the optimal solutions.


We tackle the MinMax problem on three fronts.
Firstly, when $k$ is relatively large, we propose an approximation algorithm which we show to be 3-approximate,
and which in our experiments produces good quality in the solutions.
Secondly, when $k$ is very small, say $k=2$ or $k=3$,
it would be much more convincing to provide an optimal solution.
This is due to the fact that
in the intended applications, each new server involves a significant cost, and a higher computation cost is well justified.
We assume a small parameter $k$.
For intractable problems, there may exist a parameter such that instances with small parameter values can be solved quickly.
With the assumption of a small $k$, MinMax is a candidate of parameterized problems.
We show that MinMax is in XP,
%
which is the class of parameterized problems that can be solved in time $f(k)n^{g(k)}$ for some computable functions $f$ and $g$, where $n$ is the input size and $k$ is a fixed parameter \cite{Flum2006Springer}.
To show that MinMax is in XP,
we introduce the concepts of client cost lines and potential server points. Based on their properties, we derive an optimal solution for MinMax which is efficient for small $k$ values, as illustrated by experimental results on
real datasets.
Thirdly, when $k$ is small but too costly for the basic optimal algorithm, we derive optimization strategies to
further tame the computation costs. We illustrate with experiments that our strategies work well on real city road networks even for $k$ up to 10.

This paper is organized as follows.
Section \ref{sec:probDef} contains the problem definitions.
Related works are discussed in Section \ref{sec:related}.
Section \ref{sec:approxalg} presents our approximation algorithm.
Sections \ref{sec:minmax}, \ref{sec:prelim}, and \ref{sec:first} describe our optimal solution.
Optimization strategies are discussed in section \ref{sec:enhancedMinMax}.
Section \ref{sec:exp} contains our experimental results. We conclude in Section \ref{sec:concl}.


\section{PROBLEM DEFINITION}

\label{sec:probDef}

We consider a problem setting as follows. Let $G = (V ,E)$ be an undirected graph with vertex set $V$ and edge set $E$. $C$ is a set of clients and $S$ is a set of servers. Each server or client is located on some edge in graph $G$. Each client $c$ in $C$ is associated with a positive weight $w(c)$, which represents the importance of a client. For example, if each client $c$ represents a residential area, then $w(c)$ can be the population of $c$. The distance between two points $p_{1}$ and $p_{2}$ on graph $G$ is denoted by $d(p_{1},p_{2})$.

Assume that for each client $c$ there is a unique server which is nearest to $c$
(this can be easily satisfied by shifting each client and server by a very small unique distance).
We denote the server that is nearest to client $c$ by $NN_{S}(c)$ and the distance between $c$ and its nearest server by $c.dist$, i.e., $c.dist = d(c,NN_{S}(c))$.
The cost value of $c$, denoted by $Cost_S(c)$, is equal to $w(c) \times c.dist$.

We are given a set of eligible edges $E^0 \subseteq E$.
The problem is to place a set of $k$ new servers, where each new server can be located at any point on any edge in $E^0$
 as long as there is no other server at that same point.
The introduction of $E^0$ is to model real-life scenarios where feasible locations for the new servers are typically restricted by factors such as city planning and zoning, financial considerations and property ownership.

%
We study the following optimal location query problem.

\begin{definition}[MinMax Problem]
Given a graph $G$ with a set $S$ of existing server locations,
the MinMax $k$-location query asks for a set of $k$ locations on eligible edges $E^0$ in $G$ to construct a set of $k$ new servers that minimizes the maximum cost of all clients. A set $P$
is returned where
\begin{equation}
 P = \argmin_{P} \ (\max_{c \in C}^{ } \{w(c) \times d(c,NN_{S\cup P}(c))\})
\end{equation}
\end{definition}

The minmax cost resulting from $P$ is given by
\begin{eqnarray}
\texttt{cmax}(P) = \max_{c \in C} \{w(c) \times d(c,NN_{S\cup P}(c))\}
\end{eqnarray}


\begin{example}
The following figure shows a road network $G$. Each line segment corresponds to an edge. Each dot corresponds to a vertex, server or client. There are 8 vertices, $v_1,...,v_8$, 2 servers, $s_1$, $s_2$, and 6 clients, $c_1, ..., c_6$, in this road network. $"v_{3}/c_{5}"$ means that client $c_{5}$ is on vertex $v_{3}$. The number next to each line segment is the distance between the two end points of the line segment. For example, $d(v_{5},v_{7})$ = 3, $NN_{S}(c_{4}) = s_{2}$, $c_{4}.dist = d(c_{4},s_{2})$ = 2. Let $w(c_{4})$ = 3, the cost value of $c_{4}$, $Cost_S(c_{4})$, is equal to $w(c_{4}) \times c_{4}.dist$ = $3 \times 2 = 6$.
\label{eg1:roadnetwork}
\end{example}
\hspace*{2cm}\includegraphics[width=1.5in]{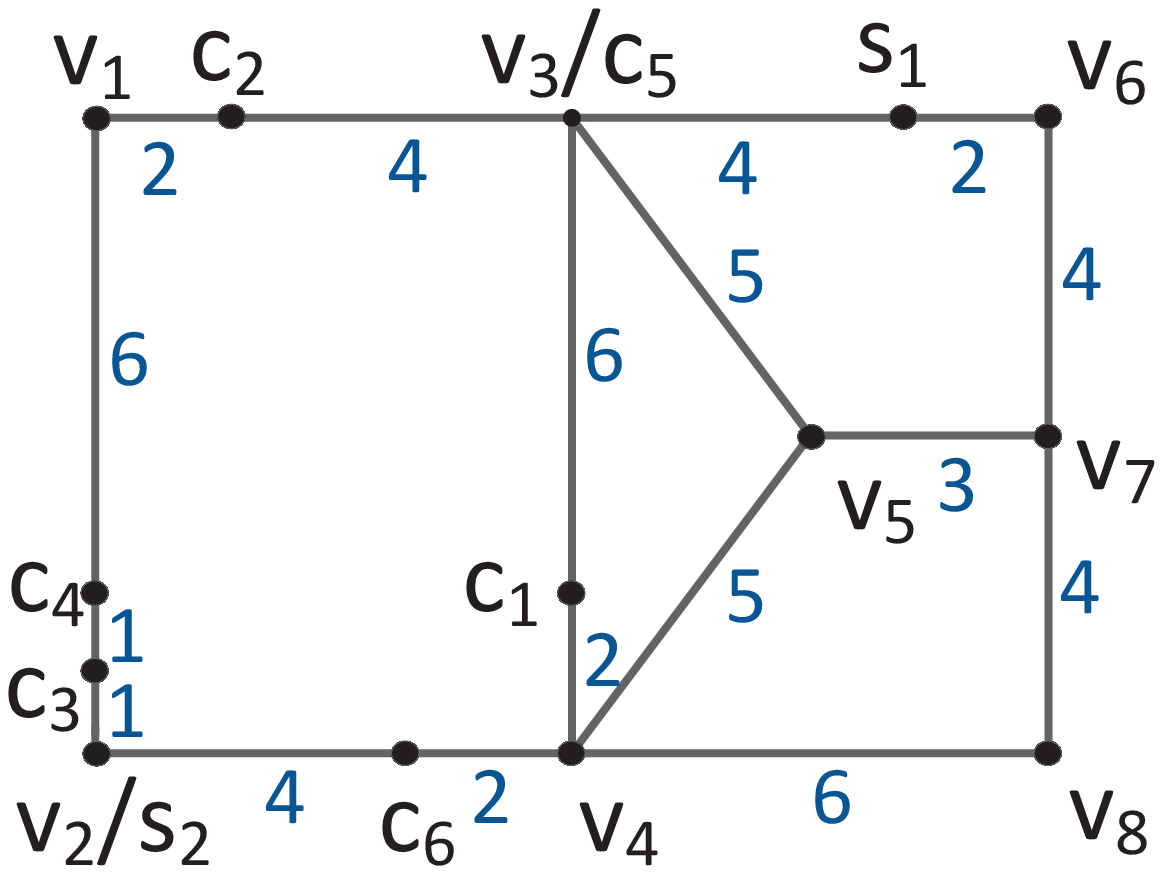}\hspace*{0mm}

Some of our notations used are listed in the following table.


\begin{center}
{\scriptsize
\begin{tabular}{|l|l|}
\hline
Notation         & Description                                                      \\ \hline
$G = (V,E)$      & road network with vertex set $V$ and edge set $E$
\\ \hline
$C$              & set of clients, $|C|=n$                                               \\ \hline
$S$              & set of servers                                                    \\ \hline
$E^0$ & set of eligible edges where new servers can be built
\\ \hline
$w(c)$           & weight of client $c$                                             \\ \hline
$NN_{S}(c)$       & server in set $S$ closest to client $c$                           \\ \hline
$c.dist$         & the distance between $c$ and its closest server $NN_S(c)$              \\ \hline
$Cost_S(c)$        & cost of client $c$, $Cost_S(c) = w(c) \times c.dist$
\\ \hline
$d(p_{1},p_{2})$ & the distance between two points $p_{1}$ and $p_{2}$ on graph $G$
\\ \hline
$Cost(c_i,x)$ &  $w(c_i) \times d(c_i,x)$
\\ \hline
$CCL_\ell(c_i)$ & Client Cost Line of client $c_i$ on $EI$ $\ell$
\\ \hline
$p.pos$ & $p=(x,y)$ is a point on a $CCL$, and $p.pos = x$
\\ \hline
$p.cost$ & $p = (x,y)$ is a point on a $CCL$ and $p.cost =y$
\\ \hline
$PSP$ & potential server point
\\ \hline
$kSP$ & a candidate set of $k$ potential server points
\\ \hline
$\texttt{cmax}(P)$ &
$\max_{c \in C} \{w(c) \times d(c,NN_{S\cup P}(c))\}$
\\ \hline
$Y.maxcost$ & $max_{p \in Y} p.cost$
\\ \hline
\end{tabular}
}
\end{center}

\section{RELATED WORK}
\label{sec:related}


We classify the related work into three types:
facility location problem in a spatial setting,
optimal location queries in a spatial setting, and optimal location queries in road networks.

\noindent
{[\it Location problems in a spatial setting]}

The facility location problem has been studied in past years \cite{drezner1995facility}. The problem is about a finite set $C$ of clients and a finite set $P$ of potential locations where new facilities can be built. The objective is to find a subset $P'$ in $P$ that optimizes a predefined cost function. It is proved that the problem is NP-hard and many studies focus on approximation algorithms \cite{Charikar99FOCS}. Surveys of location facility problems can be found in
\cite{drezner1995facility,nickel05LT,farahani09FL}.

MinMax bears some similarity to the minimax facility location problem (also called the $k$-center problem):
given a point set $X$, distance measure $d$, find a set $S \subseteq X$, $|S| = k$, so that the value of $max_{p \in X}( min_{q \in S}(d(p,q))$ is minimized. This problem is shown to be NP-hard \cite{Hochbaum85or}. For Euclidean space, an algorithm with time complexity $O(|P|^{O(\sqrt{k})})$ is known \cite{Hwang93alg}. An $(1+\epsilon)$ approximation algorithm is given in \cite{Kumar10CGA} that runs in $O(k^{|X|})$ time, and the authors show empirically that it is efficient for small $k$ values, say $k \leq 5$. The farthest point clustering method, which
greedily picks the farthest point from the selected points until $k$ centers are found, achieves 2-approximation \cite{Feder88STOC}.
An integer programming approach is taken by \cite{Calik2013} and their empirical studies consider graphs with up to 3038 vertices.

The simple plant location problem takes a set $I$
of potential sites for plants, a set of clients, costs for setting up plants at sites $i \in I$ , and transportation costs from $i \in I$ to $j \in J$ as inputs. It computes a
set $P \subseteq I$ for plant locations so that the total cost of satisfying
all client demands is minimal
\cite{krarup83jor}.
%
The online variant of facility location is studied in \cite{Meyerson01FOCS}, which assumes that the clients join the network one at a time, and the problem is to construct facilities incrementally whilst
minimizing the facility and service cost.

In general, the above problems differ from ours in that no existing servers (facilities) are given, and there is a finite set of potential locations in a $L_p$ space, whereas we assume that a road network is given with existing servers.

\noindent
{[\it OLQ in a spatial setting]}

As a variation of the facility location problem, optimal location query (OLQ) assumes that the potential facility set $P$ is the whole graph and there are some facilities built in advance. Most existing works consider $L_p$ distances. The algorithm in \cite{cardinal2006min} minimizes the maximum cost between a client to each facility when a new facility is built.
There have also been studies on the problem of finding $k$ locations to optimize the total weight of clients that become attracted by the new servers, assuming that each client is attracted to its nearest server. We refer to this problem as MaxSum. \cite{du2005optimal} solves the MaxSum location query in spatial network with three different methods. \cite{choi2012scalable} considers the MaxSum problem in a different way. Given some weighted points in spatial databases, the algorithm finds the location of a rectangular region that maximizes the sum of the weights of all the points covered by the rectangle. \cite{cabello2006reverse} studies both the MaxSum and MinMax problems in the $L_{2}$ space.

Yet another variation is the min-dist optimal location problem, which is to find a new site to minimize the average distance from each client to its closest server.
A progressive algorithm is proposed in \cite{zhang2006progressive}, which finds the exact answer given a finite number of potential locations. More recent works can be found in \cite{qi12icde,qi14www}.
The MaxBRNN problem, which finds an optimal region to build a new server that can attract the maximum number of clients in the $L_2$ space, is studied in \cite{wong2009efficient}. An improved solution is proposed in
\cite{liu2013new}.

\noindent
{\it [OLQ in road networks]}

The problems of proximity queries, including $MinMax$,
among sets of moving objects in road networks are studied in \cite{xu10sigmod}.
The MinMax location query problem in road networks with $k=1$
is first investigated in \cite{xiao2011optimal}.
Their solution adopts a divide-and-conquer paradigm.
The time complexity of their algorithm is $O(|V|+|S|+|C|)^{2}log(|V|+|S|+|C|)$  for MinMax location query.
%
%

An improved algorithm for MinMax with $k=1$ is given in \cite{chen2014efficient}.
The authors introduce a new concept called the nearest location component for each client $c$, denoted by $NLC(c,d)$, which is defined to be a set of all points on edges in $G$
at distances no more than $d$ from $c$.
The algorithm first computes the shortest distance between each client $c$ and its closest server.
Then, it sorts all clients
by their cost values.
$NLC$s are built starting from the client with the largest cost to look for a maximum integer $m$, such that
certain critical intersection of the $NLC(c_j,d_j)$,
where $d_{j} = Cost(c_{m})/w(c_{j})$ for $j \in [1,m]$, is non-empty.
The time for this step is $O(m|V|log|V|)$. An optimal location $p$ is proved to be at such an intersection,
and the time complexity to find such a location
is $O(|V||C|log|C|)$. The total complexity is $O(m|V|log|V|+|V||C|log|C|)$.
The problem becomes NP-hard when $k > 1$, an approximation algorithm is proposed in \cite{chen2014efficient}.

\section{approximation algorithm}
\label{sec:approxalg}

\begin{algorithm}[tbp]
\SetKwInOut{input}{Input}\SetKwInOut{output}{Output}
{\scriptsize
\input{$G$, $S$, $k$, eligible edges $E^0$, sorted $C$: $c_1, ..., c_n$}
\output{minmax cost: $\texttt{minmax}(G,C,S,k)$,$A$}
\Begin{
$A \gets \emptyset$\;
\For{$m = 1, ..., k$}{
    $A \gets A \cup \{ c' \}$ where $c'$ is the client with highest cost in ${c_1, ..., c_n}$.\;
    Compute $Cost_{S \cup A}(c_i)$ for each client $c_i \in c_1, ..., c_n$.\;
}
Return maximum $Cost_{S \cup A}(c_i)$ and $A$, where $1 \leq i \leq n$.\;
}
}
\caption{$AppMinMax(G,C,S,k)$}
\label{alg:CodeApprox2}
\end{algorithm}

A greedy algorithm is proposed in \cite{chen2015OLQrep}, which is to select one optimal location at a time until $k$ locations are selected. However, this can give an arbitrarily poor result. A simple example is where 2 new servers are to be built for 2 clients. The best locations are
at the 2 client sites while the greedy algorithm chooses one site in between the 2 clients.

We propose a simple algorithm AppMinMax that collects
new server locations in a set $A$, which repeatedly selects
the next server location at the location of $c_i$ with the maximum $Cost_{S \cup A}(c_i)$, until there are $k$ servers
in $A$. AppMinMax is shown in Algorithm \ref{alg:CodeApprox2}.

 \begin{theorem}
 Algorithm \ref{alg:CodeApprox2} is a 3-approximation algorithm
 for metric distances.
 \end{theorem}

{\small PROOF}:
Assume that $n$ clients $c_1$, $c_2$, $...$, $c_n$ are ordered by non-decreasing costs, i.e., $Cost_S(c_1) \geq Cost_S(c_2) \geq ...  \geq Cost_S(c_n)$. If $Q$ is an optimal solution, there must exist an integer $m$, $k+1 \leq m \leq n$, such that $Cost_S(c_{m+1}) \leq \texttt{cmax}(Q) \leq Cost_S(c_m)$. Suppose the result returned by the approximation algortihm is $A$. In a possible case, $A$ may build $k$ servers on clients $c_1$, $...$, $c_k$, so $\texttt{cmax}(A) \leq Cost_S(c_{k+1})$. $Q$ contains $k$ new servers $s'_1$, $...$, $s'_k$. 
Let $C'_i$ be the set of clients in $\{c_1, ..., c_m\}$
whose nearest server in $S \cup Q$ is $s'_i$.
The cost $Cost_{S \cup Q}(c)$ of each client $c$ in $C'_i$ is less than or equal to $\texttt{cmax}(Q)$. $C'_1 \cup ... \cup C'_k = \{c_1, ... c_m\}$. Let us take the following subgraph for
illustration:

\vspace*{-3mm}
\begin{center}
\includegraphics[width=1.6in]{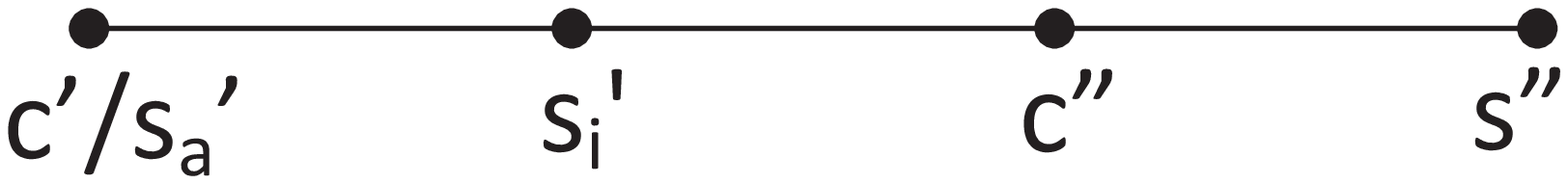}
\end{center}

In Algorithm \ref{alg:CodeApprox2}, in each iteration, we build a new server $s'_a$ on the client $c'$, where $c'$ has the highest cost. $c'$ must belong to a client set $C'_i$ because $Cost_S(c')$ cannot be smaller than $Cost_S(c_m)$. Suppose $c''$ is another client in $C'_i$ and $s'_i$ is the nearest server of both $c'$ and $c''$ in optimal solution $Q$. According to triangle inequality, $Cost(c'',s'_a) = w(c'')*(d(c'',c')) \leq w(c'')*(d(c'',s'_i)+d(s'_i,c'))$ = $w(c'')*d(c'',s'_i)+w(c'')*d(s'_i,c')$. $w(c'')*d(c'',s'_i)$ is the cost of $c''$ in the optimal solution $Q$, so $w(c'')*d(c'',s'_i) \leq \texttt{cmax}(Q)$.
Also,
if $w(c'') \leq w(c')$, $w(c'')*d(s'_i,c') \leq w(c')*d(s'_i,c') \leq \texttt{cmax}(Q)$, so $Cost(c'',s'_a) \leq 2 cmax(Q)$.



If $w(c'') > w(c')$, suppose $w(c'') = \gamma w(c')$ where $\gamma > 1$ and $s''$ is the nearest server of $c''$ in $S \cup A'$ where $A'$ is the set of locations for new servers before adding $c'$. Since in Algorithm \ref{alg:CodeApprox2}, $c'$ has the highest cost in the current iteration, $Cost_{S \cup A'}(c'') \leq Cost_{S \cup A'}(c')$. $Cost(c'',s'') = Cost_{S \cup A'}(c'') \leq Cost_{S \cup A'}(c') \leq Cost(c',s'') \leq w(c')*(d(c',s'_i)+d(s'_i,c'')+d(c'',s'')) \leq cmax(Q) + \frac{cmax(Q)}{\gamma} + \frac{Cost(c'',s'')}{\gamma}$, $Cost(c'',s'') \leq \frac{\gamma +1}{
\gamma -1} cmax(Q)$. Since $Cost(c'',s'_a) \leq w(c'')*d(c'',s'_i)+w(c'')*d(s'_i,c') = w(c'')*d(c'',s'_i)+ \gamma w(c')*d(s'_i,c') \leq cmax(Q) + \gamma cmax(Q)$, $Cost(c'',s'_a) \leq (1+\gamma) cmax(Q)$. The cost of $c''$ after building $s'_a$ is $\min(Cost(c'',s'')$, $Cost(c'',s'_a)) \leq \min(\frac{\gamma +1}{\gamma -1}, 1+\gamma)cmax(Q)$. When $\gamma > 1$, it is easy to verify that $\min(\frac{\gamma +1}{\gamma -1}, 1+\gamma) \leq 3$. So the cost of $c''$ after building $s'_a$ is smaller than or equal to $3cmax(Q)$.



In both cases, the cost of all other clients in $C'_i$ after building $s'_a$ on $c'$ will be smaller than $3cmax(Q)$.
After adding $c'$ to $A$ and recomputing the cost of all clients in $C$, if the client with the highest cost is in set $C'_i$, that means all clients have a cost smaller than $3cmax(Q)$. Otherwise, the chosen client should belong to another set $C'_j$ where $1 \le j \le k$ and the cost of all clients in $C'_j$ will be reduced to less than $3cmax(Q)$ with a new server. Since $C'_1 \cup ... \cup C'_k = \{c_1, ... c_m\}$, within $k$ iterations, we can put at least one server in each client set if the client set contains a client with the highest cost. The maximum cost returned by Algorithm \ref{alg:CodeApprox2} must be smaller than $3cmax(Q)$ after $k$ iterations, we conclude that Algorithm \ref{alg:CodeApprox2} is a 3-approximation algorithm.
\done

\begin{figure*}[htbp] \centering{
\includegraphics[width=1.4in]{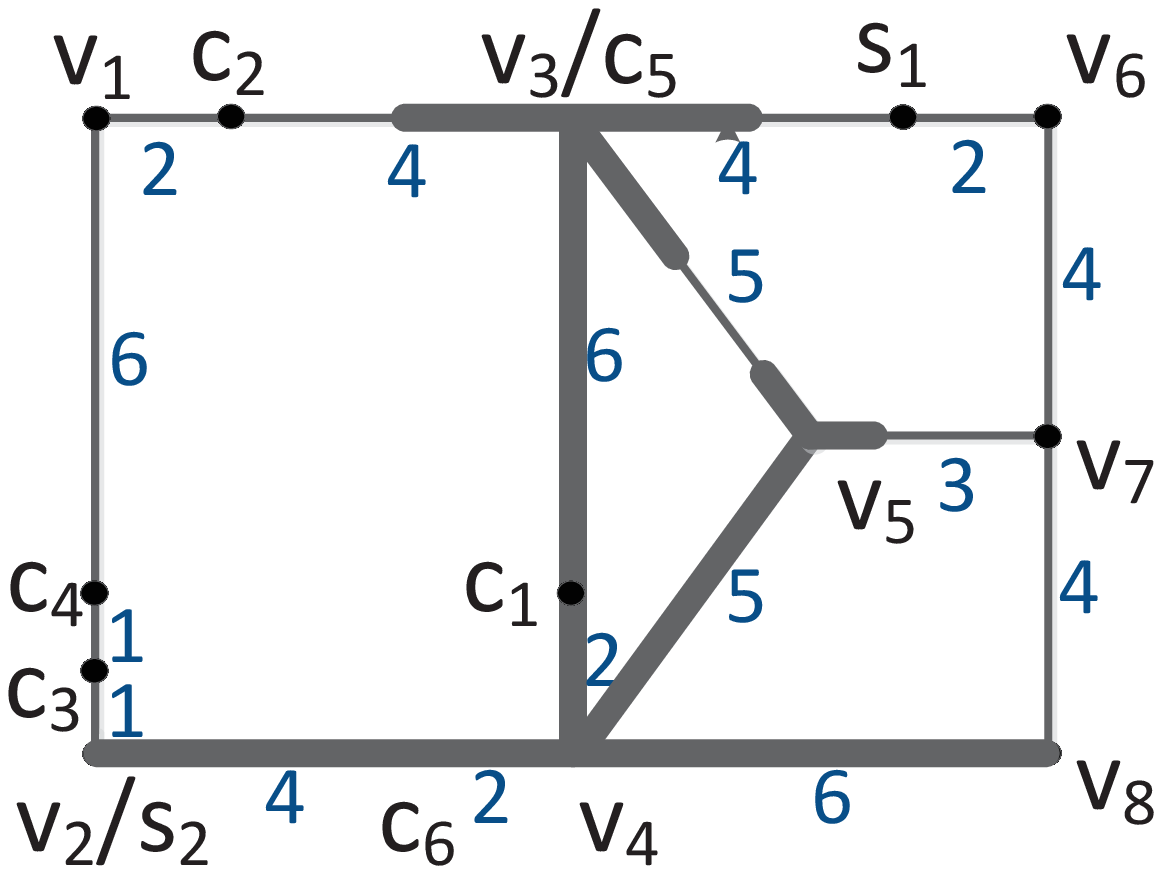}
\hspace*{-3mm}
\includegraphics[width=1.4in]{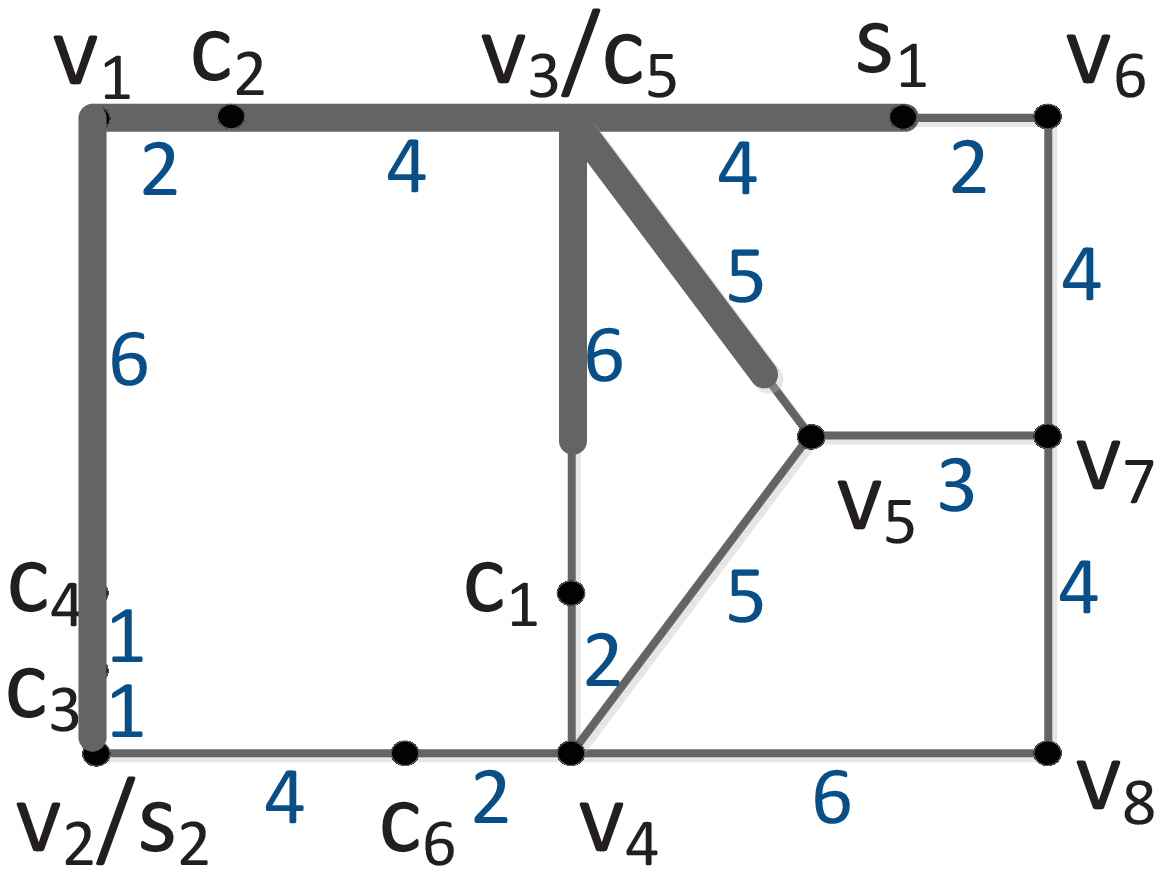}
\hspace*{-3mm}
\includegraphics[width=1.4in]{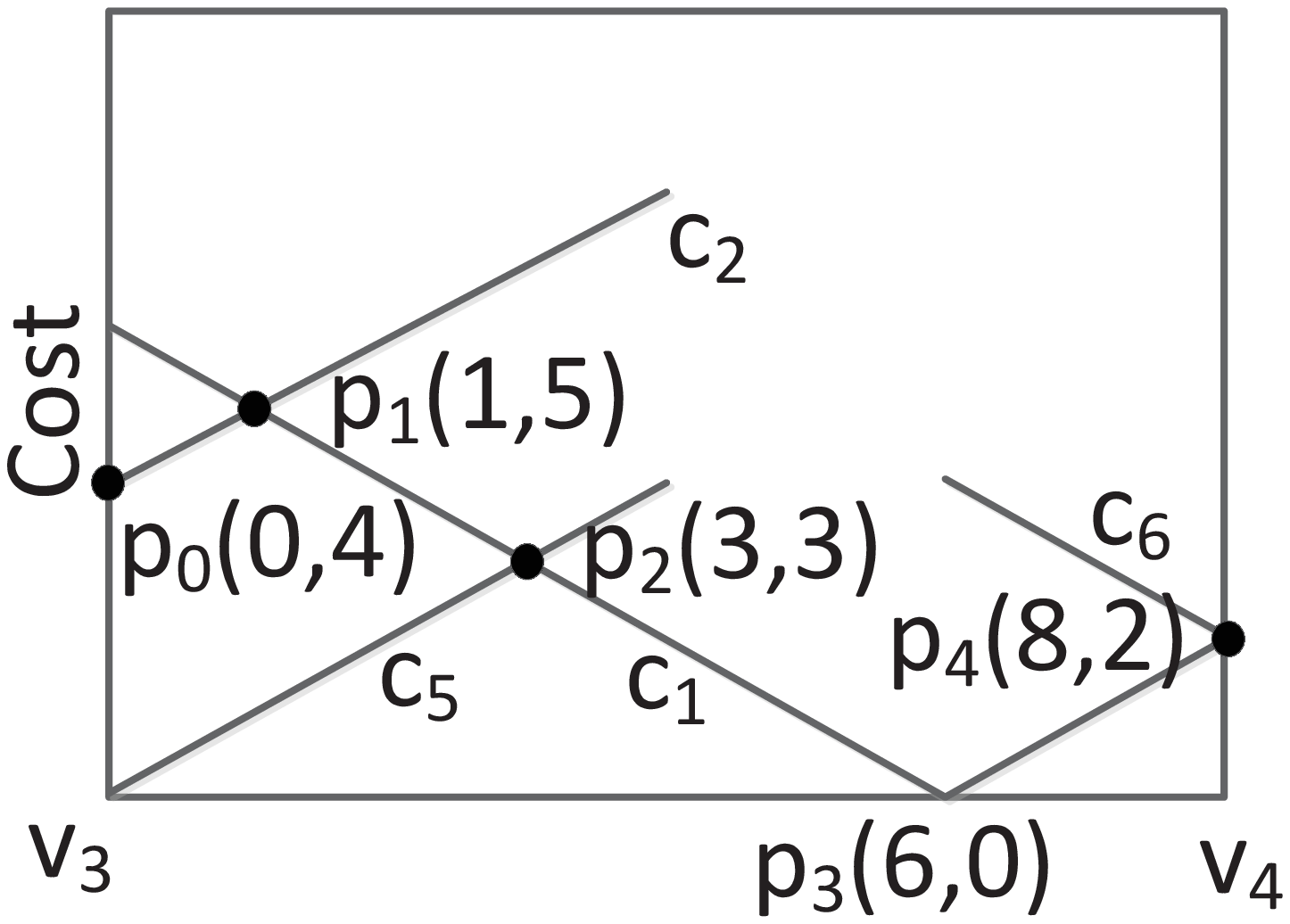}
\hspace*{-3mm}
\includegraphics[width=1.4in]{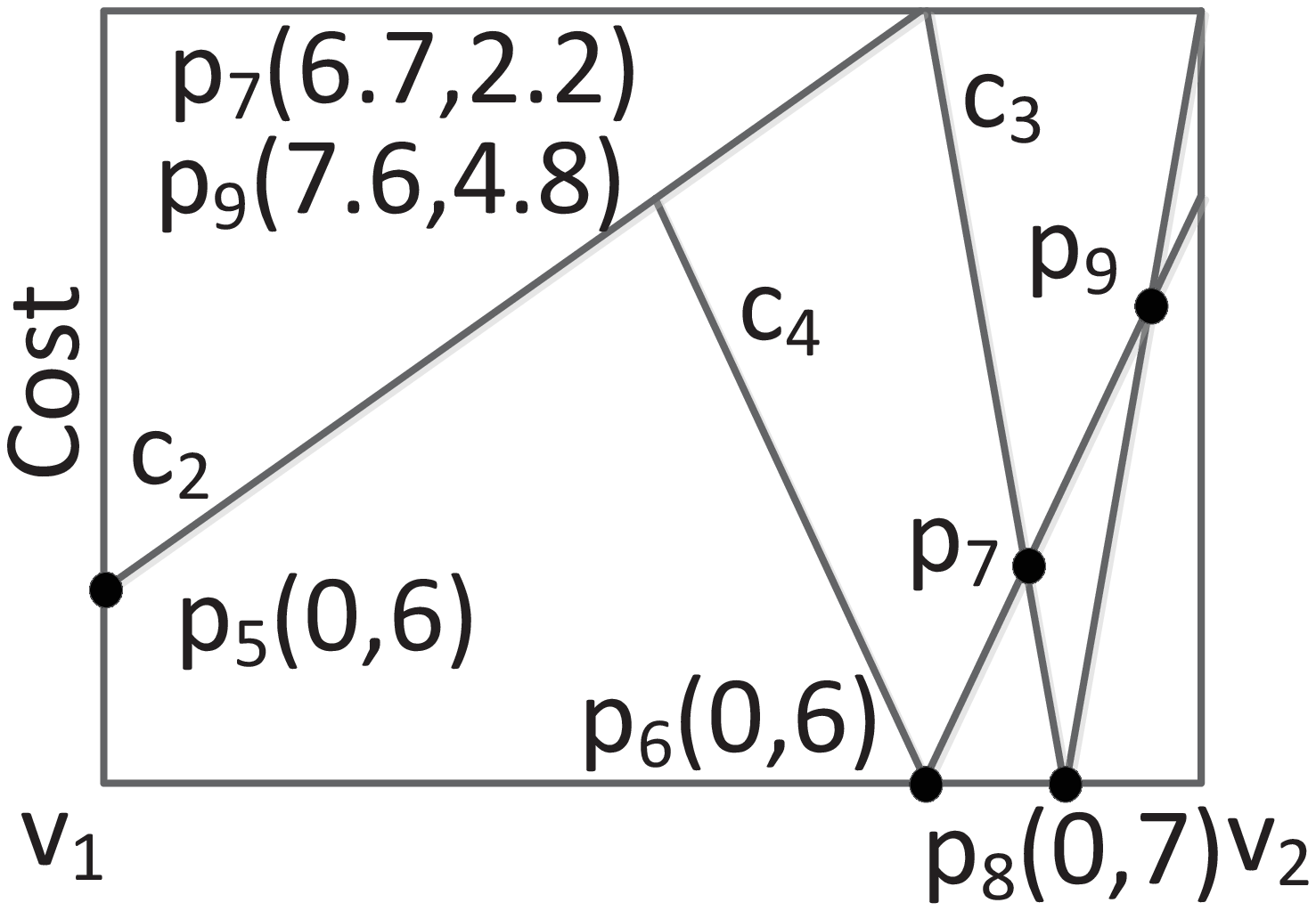} \hspace*{-3mm}
\includegraphics[width=1.45in]{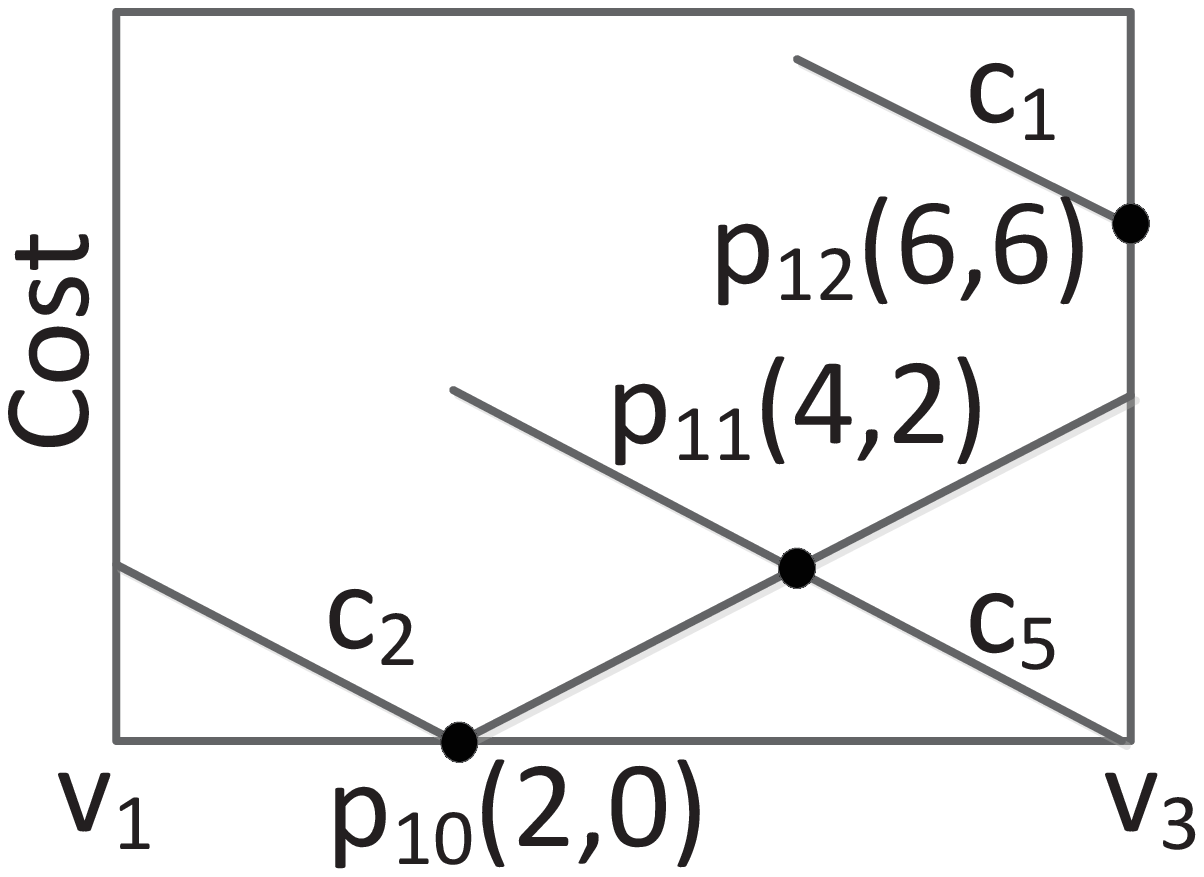} 
\\
(a)
\hspace*{30mm}(b)
\hspace*{30mm}(c)
\hspace*{30mm}(d)
\hspace*{30mm}(e)
\caption{(a) Graph $G$, $w(c_1)$=$w(c_2)$=$w(c_5)$=$w(c_6)$=1, $w(c_3)$=8, $w(c_4)$=3, $NLC(c_{1})$ in bold ; 
(b) $NLC(c_2)$ in bold; (c) $CCLs$ for $G$: plotting $Cost(c_i,p)$ on edge $[v_3,v_4]$;  $p_0, p_1, p_2, p_3$ and $p_4$ are $PSP$s ; (d) $CCLs$ for $G$: plotting $Cost(c_i,p)$ on edge $[v_1,v_2]$;  $p_5, p_6, p_7, p_8$ and $p_9$ are $PSP$s ; (e) $CCLs$ for $G$: plotting $Cost(c_i,p)$ on edge $[v_1,v_3]$;  $p_{10}$, $p_{11}$, and $p_{12}$ are $PSP$s}
\label{figG1}
}
\end{figure*}

\smallskip

\begin{lemma}
Algorithm \ref{alg:CodeApprox2} is not ($3-\epsilon$)-approximate for $\epsilon > 0$.
\end{lemma}

\begin{figure}[h!]
\begin{center}
\hspace*{-3mm}
\includegraphics[width=1.5in]{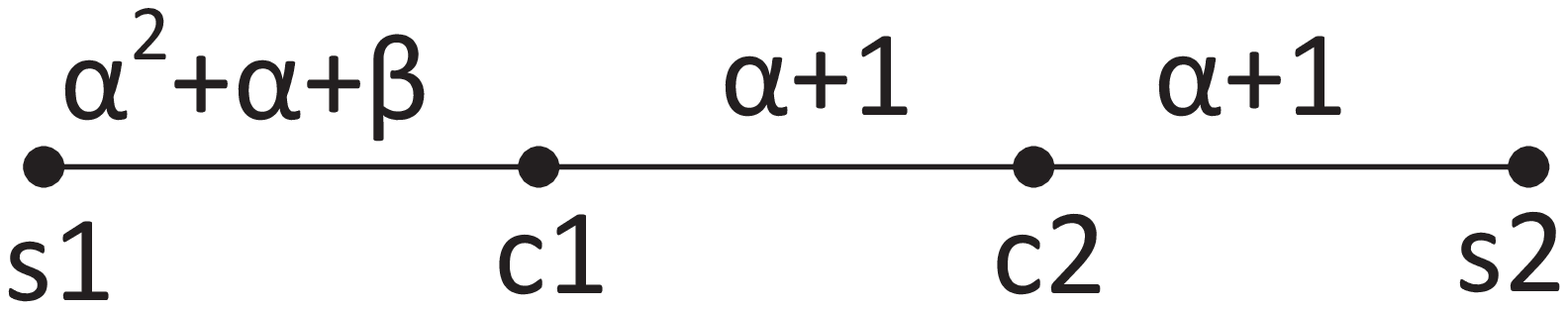}
\includegraphics[width=1.5in]{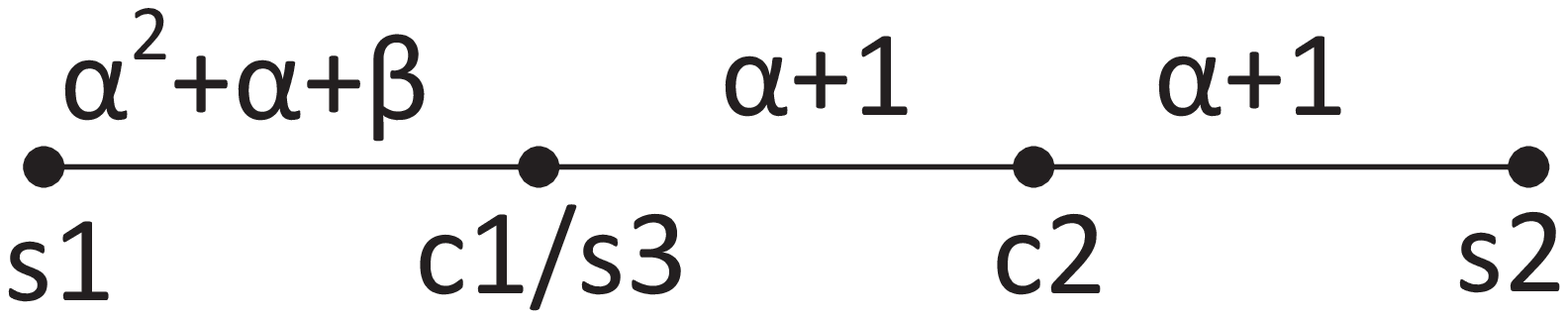}
\\
(a) \hspace*{1.3in} (b)
\end{center}
\vspace*{-3mm}
\caption{(a) A road network with $w(c_1)$=1, $w(c_2)$ = $\alpha$, $\alpha > 0$ and $\beta > 0$; (b) Solution of the greedy algorithm when $k$ = 1}
\vspace{-2mm}
\label{fig3}
\end{figure}

Figure \ref{fig3} (a) shows a road network with 2 existing servers, $s_1$, $s_2$, and 2 clients, $c_1, c_2$. $d(s_{1},c_{1})$ = $\alpha^2+\alpha+\beta$, $d(c_{1},c_{2})$ = $\alpha+1$ and $d(c_{2},s_{2})$ = $\alpha+1$ where $\alpha > 0$ and $\beta > 0$. Assume that the client weights are the following: $w(c_1) = 1, w(c_2) = \alpha$. $c_1$ and $c_2$ have their nearest servers as $s_1$ and $s_2$, respectively. Since the nearest server of $c_1$ is $s_1$, $d(s_{1},c_{1})$ = $\alpha^2+\alpha+\beta < 2\alpha + 2$ = $d(c_{1},s_{2})$. We can get $\alpha < \frac{\sqrt{9-4\beta}+1}{2} < 2$. $Cost_S(c_1)$ = $w(c_1) \times c_1.dist$ = $\alpha^2+\alpha+\beta$ and $Cost_S(c_2)$ = $\alpha^2+\alpha$. The client ordering is $c_1$, $c_2$.

Figure \ref{fig3} (b) shows the case when $k=1$.  Algorithm \ref{alg:CodeApprox2} builds a new server $s_3$ at $c_1$. The nearest server of $c_1$ is $s_3$ and $Cost_S(c_1)$ = 0. The maximum cost of all client is $cmax(A)$ = $Cost_{S \cup A}(c_2)$ = $\alpha^2+\alpha$. In an optimal solution, if we build the new server between $c1$ and $c2$ and the distance to $c_1$ is $\alpha$, then $Cost_S(c_1) = Cost_S(c_2) =  \alpha$ and the maximum cost among all clients is $cmax(Q)$ = $\alpha$. So, Algorithm \ref{alg:CodeApprox2} is not ($3-\epsilon$)-approximate for $\epsilon > 0$ when clients have different weights, since $cmax(A)$/$cmax(Q)$ = $\alpha +1$ and $0<\alpha<2$, and we can set $\alpha > 2 - \epsilon$.
\done

We have a better guarantee when clients are unweighted.

\begin{lemma}
Algorithm \ref{alg:CodeApprox2} is 2-approximate when all clients have unit weight.
\end{lemma}

\smallskip

{\small PROOF}:
Assume that $n$ clients $c_1$, $c_2$, $...$, $c_n$ are ordered by non-decreasing costs, i.e., $Cost_S(c_1) \geq Cost_S(c_2) \geq ...  \geq Cost_S(c_n)$. Each client has unit weight. If $Q$ is an optimal solution, there must exist an integer $m$, $k+1 \leq m \leq n$, such that $Cost_S(c_{m+1}) \leq \texttt{cmax}(Q) \leq Cost_S(c_m)$. Suppose the result returned by the approximation algorithm is $A$. In a possible case, $A$ consists of $k$ servers at the clients $c_1$, $...$, $c_k$, so $\texttt{cmax}(A) \leq Cost_S(c_{k+1})$.


Optimal solution $Q$ contains $k$ new servers $s'_1$, $...$, $s'_k$. Each new server $s'_i \in Q$ is the nearest server of a set of clients $C'_i$ where $C'_i \subset \{c_1, ... c_m\}$ for $ 1 \leq i \leq k$ and the cost of each client in $C'_i$ is less than or equal to $\texttt{cmax}(Q)$. $C'_1 \cup ... \cup C'_k = \{c_1, ... c_m\}$. In Algorithm \ref{alg:CodeApprox2}, for each iteration we build a new server $s'_a$ on a client $c'$, where $c'$ must belong to a client set $C'_i$. Since each client in $C'_i$ has unit weight and they share the same nearest server $s'_i$ with maximum cost $\texttt{cmax}(Q)$, according to triangle inequality, the cost of all other clients in $C'_i$ after building $s'_a$ will be no more than $\texttt{cmax}(Q)+\texttt{cmax}(Q) = 2\texttt{cmax}(Q)$.

 After adding $c'$ to $A$ and recomputing the cost of all clients in $C$,
if the client with the highest cost is in set $C'_i$, that means each client has a cost of no more than $2cmax(Q)$. Otherwise, the client chosen should belong to another set $C'_j$ where $1 \le j \le k$ and the cost of all clients in $C'_j$ will be reduced to no more than $2cmax(Q)$ with a new server.
After $k$ iterations, the maximum cost returned by Algorithm \ref{alg:CodeApprox2} must be no more than $2\texttt{cmax}(Q)$, so $\texttt{cmax}(A) \leq 2\texttt{cmax}(Q)$, which proves that Algorithm \ref{alg:CodeApprox2} is a 2-approximation algorithm when all clients have unit weight.
\done

Consider the time complexity of Algorithm \ref{alg:CodeApprox2}. Computing the distance from each client to its nearest server takes
$O(|V|\log |V| + |C|)$ time,
as shown in \cite{chen2015OLQrep}. Sorting takes $O(|C|log|C|)$ time. After building each new server, it takes $O(|V|log|V|)$ time to recompute all client costs.
Thus, the time complexity is $O(|C|log|C|+ k|V|log|V|)$. The storage space for running Dijkstra’s algorithm is $O(|V|)$ and the space for storing costs of all clients is $O(|C|)$ . The storage complexity is $O(|V|+|C|)$.

\section{Exact Solution Framework}
\label{sec:minmax}

Our exact solution will make use of the \textbf{Nearest Location Components} as defined in \cite{chen2014efficient}:
$NLC(c) = \{p|d(c,p) \leq c.dist $ and $p$ is
a point on an edge in $G \}$.
For each client $c$, $NLC(c)$ is a set of all points on edges in $G$ with a distance to $c$ of at most $c.dist$. In Figure \ref{figG1} (a), the bold line segments correspond to $NLC(c_{1})$, where the distance between each point in $NLC(c_{1})$ to $c_{1}$ is at most $c_{1}.dist = d(c_{1},s_{2})$ = 8.

For clarity, we shall overload the symbol $P$ to stand for both a set of locations and the set of servers built on these locations, and $c_i$ to also stand for the location of $c_i$.

The MinMax problem is to find a set $P$ of $k$ locations on the road network for new servers so that with the new and existing servers, the maximum among the costs $Cost_{S \cup P}(c)$ of all clients $c \in C$ is minimized.
%
Let us name the $n$ clients $c_1, c_2, ..., c_n$, so that the clients are ordered by non-increasing costs, i.e., $Cost_S(c_1) \geq Cost_S(c_2) \geq ...  \geq Cost_S(c_n)$.

\label{sec:framework}

Our solution framework is based on 2 main observations:
(O1) From the definition of $NLC$, it is clear that in order to reduce the cost of a client $c$, a new server must be located on $NLC(c)$.
(O2) Since the clients are sorted by $Cost_S(c)$, if $Q$ is an optimal solution, there exists $m$, $k \leq m \leq n$, such that $Cost_S(c_{m+1}) \leq \texttt{cmax}(Q) \leq Cost_S(c_m)$.

From (O1), promising locations for new servers are on the $NLC$s of clients with greater costs.
From (O2), it suffices to consider $NLC$s for client sets of $C_i = \{c_1, ..., c_i\}$ with increasing $i$, and to stop at $C_m$ when the best solution $P$ for $C_m$ introduces a cost $Cost_{S \cup P}(c_j)$, for some $1 \leq j \leq m$, higher than $Cost_S(c_{m+1})$.
This suggests the following iterative process:

\begin{algorithm}[h!]
\SetKwInOut{input}{Input}\SetKwInOut{output}{Output}
{\small
\input{$G$, $S$, $k$, $E^0$, sorted client set $C$: $c_1, ..., c_n$}
\output{minmax cost and $k$ optimal locations}
\Begin{
\For{client set $C_i = \{c_1, ..., c_i\}$, $ i \in k..n$}{
    \textbf{Step 1}: On each edge, find all locations for new servers which have the potential to be in the optimal solution\;
    \textbf{Step 2}: Examine each combination $P$ of $k$ or less locations derived in Step 1.
    Find the combination $P$ with the minimum $\texttt{cmax}(P)$ value assuming that the client set is $C_i$. Break if
    this minmax value is greater than $Cost_S(c_{i+1})$\;
    }
Return the best $k$ potential locations found and the corresponding minmax cost\;
    }
}
\caption{Solution Framework}
\label{alg:framework}
\end{algorithm} 
%
%
%
%

For Algorithm 2, a naive procedure for Step 1 is the following:
on each edge, for each subset $x$ of $C_i$, find a potential location for a new server to minimize the maximum cost of clients in $x$.
%
Let us consider the road network in Example \ref{eg1:roadnetwork}.
Suppose $k=2$; we need to build 2 new servers.
The clients $c_i$ are sorted in the order of $Cost_S(c_i)$.
$c_1, c_2, c_3$ have the highest cost of 8, and we first
try to reduce the costs of $C_2 = \{c_1, c_2\}$ with the new servers.
$NLC(c_1)$ and $NLC(c_2)$ are shown in Figures \ref{figG1}(a)
and \ref{figG1}(b).
The two $NLC$s overlap
on edges $[v_1, v_3]$, $[v_3, v_6]$,
$[v_3, v_4]$ and $[v_3, v_5]$.
An optimal location among these would be a point on
edge $[v_3, v_4]$ with the same
distance from $c_1$ and $c_2$.
A new server built
at such a location in this segment will
reduce the costs of both $c_1$ and $c_2$ to the same value.
However, since we have two servers, we may also consider the best
locations to simply reduce the cost of $c_1$ by a new server,
and the best locations to reduce the cost of $c_2$ by
another new server.
Thus, all possible subsets of $C_2$ are considered on each edge.
Step 2 in the solution framework will then pick the best combination of $k$ locations to
be the choice for $C_i$.

The two steps are repeated with $C_3 = \{c_1, c_2, c_3\}$.
For reducing the cost of $c_3$, we may consider segments on
$[v_1, v_2]$.
The next highest cost client is $c_4$, with $Cost_S(c_4) = 6$.
If the best possible reduced costs for $c_1, c_2, c_3$ are smaller than 6,
then we also try to reduce the cost of $c_4$ by examining $NLC(c_4)$.
This continues until the current minmax cost is higher than the cost
of the next highest cost client.

However, the above naive procedure for Step 1 is costly.
If there are $\gamma$ client $NLC$s overlapping with the edge $\ell$, the number of subsets will be $O(2^\gamma)
\subseteq O(2^n)$. The difficulty lies with how to avoid this exponential factor.
Based on the concepts of client cost lines and potential server points, to be introduced in Section \ref{sec:prelim}, we are able to reduce the complexity of Step 1 to $O(\gamma^2)$ $\subseteq O(n^2)$ for each edge, as the number of potential locations is limited to $O((\gamma^2))$. 
Such a solution is presented in
Section \ref{sec:first}.

\section{CCL and PSP}
\label{sec:prelim}

In this section we study the issues in Step 1 of the solution framework. We introduce the notions of client cost lines ($CCL$) and potential server points ($PSP$), and derive some useful properties concerning $CCL$s and $PSP$s.

\subsection{Client Cost Lines ($CCL$)}

Let $\ell = [a,b]$ be an edge covered by $NLC(c_i)$.
Call $a$ and $b$ the boundaries of $\ell$.
%
Let $x$ be a point in $\ell$ such that $x \in NLC(c_{i})$. Define $Cost(c_i,x) = w(c_i) \times d(c_i,x)$.
Given client $c_i$, define function
$f_i(x) = Cost(c_i,x)$ for $x \in NLC(c_i) \cap \ell$, which is a piecewise linear function of $x$.
Note that $f_i(x)$ is undefined for $x \not\in NLC(c_i)$.
Call the plotted line for function $f_i$ on
the positions on $\ell$ the \textbf{Client Cost Line} for $c_i$ on $\ell$, denoted by $CCL_\ell(c_i)$.
Hence, given a point $p=(x,y)$ on $CCL_\ell(c_i)$,
$y = Cost(c_i,x)$.
We also denote $x$ by \textbf{\it \textbf{p.pos}} and $y$ by {\it \textbf{p.cost}}.


\begin{example}[$CCL$]
Figure \ref{figG1}(c) shows the $CCL_\ell(c_i)$ for clients $c_1,c_2,c_5,c_6$ on the edge $\ell=[v_3, v_4]$. The $x$-axis is for positions on $\ell$, and the
$y$-axis is for the costs.
Figures \ref{figG1}(d) and \ref{figG1}(e) show $CCL$s on edges $\ell = [v_1, v_2]$ and $\ell = [v_1, v_3]$.
\end{example}


The union of $CCL_\ell(c_i)$ for all $\ell$ is denoted by {\it \textbf{$CCL(c_{i})$}}.
The values of $p.pos$ of points on $CCL_\ell(c)$
correspond to the projection of the $CCL$ on $\ell$.
The union of the projections of
$CCL_\ell(c_i)$ for all edges $\ell$ spans the set of possible positions for a new server to attract $c_i$.
Given a point $p=(x,y)$ on $CCL_\ell(c_i)$, if
$x$ is the location of a new server, then $y$ is a possible new minmax cost.
For convenience, we also say that a new server is placed at $p$ when it is placed at $x$.

$CCL_\ell(c_i)$ consists of one or two linear pieces, which may be connected or disconnected.
$CCL_\ell(c_i)$ \textbf{ends at a point} $(x,z)$ if $f_i$ is undefined in a non-empty interval $(x,x']$ or $[x',x)$.
%
For example, in Figure \ref{figG1}(c) , $\ell = [v_3,v_4]$, and
$CCL_\ell(c_6)$ ends at the
 point $(6,4)$.
If $CCL_\ell(c_i)$ ends at $(x,z)$, then $z = Cost_S(c_i)$, and $z$ is the highest cost value for $CCL(c_i)$.

%
If a new server is placed at a point on $CCL_\ell(c_i)$ other than an end point,
the cost for $c_i$ is reduced as $c_i.dist$ becomes smaller.
%
The following definition helps to describe how a point is related to reduction in client costs.


\begin{definition}[$c_i$ covered by $p$]
We say that a client $c_i$ is \textbf{covered} by a point $p=(x,\lambda)$ on a $CCL$ on edge $\ell$ if given a new server $s$ at $x$ on $\ell$, and $S' = S \cup \{s\}$,
$Cost_{S'}(c_i) \leq \lambda$.
\end{definition}


Note that $p$ does not need to be in $CCL_\ell(c_i)$ in the above definition.
For example, in Figure \ref{figG1}(c), $p_1$ covers $c_6$, since $Cost_S(c_6)$ is less than the cost at $p_1$. $p_1$ also covers $c_1$ and $c_2$, being the intersection of $CCL_\ell(c_1)$ and $CCL_\ell(c_2)$.
%
The idea of covering is to identify subsets of clients that do not need to be covered by other new servers. In the above, $c_6$ is ``covered'' by $p_1$ because if a server is built at $p_1.pos$ attaining a cost of $p_1.cost$ for $c_1$ and $c_2$, then we need not try to cover $c_6$ by another server, since reducing the cost of $c_6$ does not reduce the overall minmax cost.
%
In Figure \ref{figG1}(c), to cover all the clients $c_1, c_2, c_5, c_6$ by a new server on $\ell$, the best position for the new server is that of $p_1$,
and the maximum $Cost(c_i,x)$ of the above clients will be minimized within $\ell$ when $x=p_1.pos$.
However, with multiple new servers, $c_1$ may be covered by a new server on another edge, and
if we only need to cover
$c_2, c_5, c_6$ by a new server on $\ell$,
the best position is that of $p_0$ instead.



Next, we introduce a notion for the effect of choosing a location $p$ on the minmax cost concerning a client set $C'$.


\begin{definition}[$zmax(C',p)$]
Given point $p$ on $CCL_\ell(c_i)$ for $c_i \in C'$ and edge $\ell$ in $G$. Let $C'' \subseteq C'$ be the set of clients in $C'$ not covered by $p$. Define $Cost_S(\emptyset)=0$ and
if $C'' \neq \emptyset$,
$Cost_S(C'') = \max(Cost_S(c) | c \in C'' )$.
$\texttt{zmax}(C',p)$ is defined to be $\max( p.cost, Cost_S(C'') )$.
\end{definition}

%

\begin{example}[$zmax(C',p)$]
To reduce the overall minmax cost we need to reduce the costs of some clients.
Consider
Figure \ref{figG1}(c).
Suppose we aim to reduce the costs of $c_1$ and $c_5$ by placing a new server on $[v_3,v_4]$.
Compare the choices of $p_1$ and $p_2$, both of which cover $c_1$ and $c_5$.
Since $p_1.cost > p_2.cost$, $p_2$ is a better choice.
We have
$\texttt{zmax}(\{c_1,c_5\},p_1)= p_1.cost$ $>$
$\texttt{zmax}(\{c_1,c_5\},p_2) = p_2.cost$.
\end{example}

%

Next, we consider some useful properties concerning $\texttt{zmax}(C',p)$. 
For a set $P$ of locations in $G$, let us
define the term $\texttt{cmax}(P) = \max_{c \in C} \{w(c)*d(c,NN_{S\cup P}(c))\}$.
A set $P$ with $|P| = k$ and minimum $\texttt{cmax}(P)$ is an optimal solution.

We map locations of a candidate solution to points of the $CCL$s.
Consider a set $P$= $\{x_1, ..., x_k\}$ of $k$ locations on $E^0$.
Each $x_i$ intersects the $CCL$ projections of a set of clients $C_i$.
For a client $c$ in $C'=C_1 \cup ... \cup C_k$,
compare the costs, $Cost(c,x_i)$, $x_i$ in $x_1,...,x_k$, and pick $x_j$ with the smallest $Cost(c,x_j)$.
We say that $c$ is \textbf{assigned to} $x_j$.
%
In Figure \ref{figG1}(c), if $k=2$, and $x_1 = v_3$,
$x_2=v_4$, then $c_2, c_5$ will be assigned to $v_3$ and
$c_1, c_6$ will be assigned to $v_4$.

For each $x_i$ in $P$, among all assigned clients, pick $c'$ whose $CCL$ has a highest cost at $x_i$.
The point of $CCL(c')$ at $x_i$ is called the \textbf{apex point} of $x_i$, denoted by $a_i$. E.g., suppose $x_i = v_3$ in Figure \ref{figG1}(c), if clients $c_2, c_5$ are assigned to $v_3$, then $p_0$ is the apex point $a_i$ of $x_i$.

The candidate solution $P$ can thus be also defined by the $k$ apex points
$a_1, a_2, ..., a_k$. Let $A = \{a_1,...,a_k\}$.
We say that $P$ is \textbf{mapped to} $A$.
Define $\texttt{amax}(A) = \texttt{cmax}(P)$.
We prove the following in Appendix 1.


\begin{lemma}
Let $P$ be a set of $k$ locations which is mapped to a set $A$ of $k$ $CCL$ points.
Let $p \in A$ and $C'$ be the set of clients covered by $p$. Let $q \not\in A$ be another point on a $CCL$, where
$\texttt{zmax}(C',p) \geq \texttt{zmax}(C',q)$.
Let $A' = A - \{p\} \cup \{q\}$,
then $\texttt{amax}(A')$ $\leq \texttt{amax}(A)$.
\label{lem:minmax}
\end{lemma}

Lemma \ref{lem:minmax} states that we can replace $CCL$ point $p$ by $q$ in the candidate solution $P$ without increasing the overall minmax cost. We shall make use of this lemma in minimizing the number of potential locations for new servers
in the next subsection.

\subsection{Potential Server Points ($PSP$)}

In this subsection, we study how to determine potential points on the $CCL$s for placing new servers.
Let us call such critical $CCL$ points the \textbf{potential server points} ($PSP$s). Denote the set of $PSP$s for $\ell$ by $PSP(\ell)$.
Our task is to determine $PSP(\ell)$ for each $\ell$.

Consider an edge $\ell = [a,b]$.
We define the \textbf{lowest boundary (LB) point} of
$CCL_\ell(c)$ as follows:
If client $c$ lies on $\ell$, thus,
$(c,0) \in CCL_\ell(c)$, then $(c,0)$ is the LB point of $CCL_\ell(c)$;
otherwise, there are two cases:
(1) if both points $(a,y_a)$ and $(b,y_b)$ are on $CCL_\ell(c)$,
and $y_a < y_b$ ($y_a > y_b$), then $(a,y_a)$ ($(b,y_b)$) is the LB point of $CCL_\ell(c)$;
(2) else, if only point $(a,y_a)$ ($(b,y_b)$) is in $CCL_\ell(c)$,
then $(a,y_a)$ ($(b,y_b)$) is the LB point of $CCL_\ell(c)$.
We say that the LB point is on $\ell$.


\begin{example}[LB point]
In Figure \ref{figG1}(c), $p_0$ is the LB point of $CCL_\ell(c_2)$, where $\ell = [v_3, v_4]$, and
$p_3$ is the LB point of $c_1$.
In Figure \ref{figG1}(d), $p_5$ is the LB point of $CCL_\ell(c_2)$, where $\ell = [v_1, v_5]$, and
$p_6$ is the LB point of $c_4$.
\end{example}


We shall show that we can restrict the selection of locations for new servers on an edge $\ell$ to the following two rules:


\begin{enumerate}
\item[\bf\texttt{R1}]
choose an intersection point of 2 or more $CCL$s.
\item[\bf\texttt{R2}]
choose a lowest boundary point of a $CCL_\ell(c)$.
\end{enumerate}




Let us denote the set of intersection points of $CCL_\ell$s by $I(\ell)$ and the set of LB points on $\ell$ by $B(\ell)$. Our first attempt
is to set the potential locations based on
$I(\ell) \cup B(\ell)$.
Let $I = \cup_{\ell} I(\ell)$ and $B = \cup_{\ell} B(\ell)$.



\begin{lemma}
A solution for MinMax exists where the $k$ new servers are located at the positions of $k$ points in $I \cup B$.
\label{lem:PSP1}
\end{lemma}

In Figure \ref{figG1}(c), $\ell = [v_3,v_4]$, $I(\ell)=\{p_1,p_2\}$, and $B(\ell) = \{p_0,p_3,p_4\}$. $PSP(\ell)$ is set to $\{p_0,p_1,p_2,p_3,p_4\}$.
A proof of the above lemma
is given in the Appendix.
We can further restrict the selection of locations for new servers on an edge $\ell$ by replacing Rule $\texttt{R1}$ with Rule
$\texttt{R3}$ in the following.

\begin{enumerate}
\item[\bf\texttt{R3}]
choose an intersection point of 2 $CCL$ linear pieces where the slope of one piece is positive and the other is negative.
\end{enumerate}





\begin{example}[Rule R3]
In Figure \ref{figG1}(d),
there are two intersection points for $CCL_\ell(c_3)$ and
$CCL_\ell(c_4)$, namely, $p_7$ and $p_9$.
However, $p_9$ is the intersection of two linear pieces both having a positive slope.
In contrast, $p_7$ satisfies Rule \texttt{R3}.
\end{example}

\begin{lemma}
A MinMax solution exists where the $k$ new servers are located at points selected by \texttt{R2} or \texttt{R3}.
\label{lem:PSP2}
\end{lemma}


{\small PROOF}:
Based on Lemma \ref{lem:PSP1}, we only need to exclude the points in $I$ which are intersections of two linear pieces with both positive or both negative slopes.
Without loss of generality, consider a point $p=(x,y)$ in $I$ on the intersection of $CCL_\ell(c)$ and $CCL_\ell(c')$, the slopes of which at $p$ are both negative, and $\ell = [a,b]$, also $Cost(c,z) > Cost(c',z)$ for $z > y$.
With similar arguments as in the proof of Lemma \ref{lem:PSP1}, we can find a point $q$
on $CCL_\ell(c)$ which is
in $I(\ell) \cup B(\ell)$ such that for the set $C'$ of clients covered by
$p$, $\texttt{zmax}(C',p) \geq \texttt{zmax}(C',q)$.
\done


Now, we have a definition for $PSP$s.

\begin{definition}
A potential server point ($PSP$) is a point selected by Rule
\texttt{R2} or Rule \texttt{R3}.
\end{definition}


\begin{lemma}
There are $O(m^2)$ potential server points on an edge which overlap with the $NLC$s of $m$ clients.
\end{lemma}


{\small PROOF}:
Each client contributes at most two linear pieces of $CCL$s on an edge $\ell$.
From Lemma \ref{lem:PSP2}, a $PSP$ is either an intersection point of $CCL$s or a lowest boundary point; the lemma thus follows. \done

\section{MinMax Algorithm}
\label{sec:first}

Here, we derive an algorithm for MinMax based on the findings above.
%
%
%
\begin{algorithm}[tbp]
\SetKwInOut{input}{Input}\SetKwInOut{output}{Output}
{\scriptsize
\input{$G$, $S$, $k$, eligible edges $E^0$, sorted $C$: $c_1, ..., c_n$}
\output{minmax cost: $\texttt{minmax}(G,C,S,k)$,$kSP$}
\Begin{
$P \gets \emptyset$\;
\For{$m = 1, ..., k$}{
     build $CCL(c_m)$ on each edge in $E^0$\;
}
Compute set $Q$ of $PSP$s and clients covered by each $PSP$ on $E^0$\;
$P \gets P \cup \{Q\}$\;
\For{$m = k+1, ..., n$}{
	$curmax \gets Cost_S(c_{m})$\;
    build $CCL(c_m)$ on edges in $E^0$\;
    remove $PSP$s in $P$ with cost $\geq curmax$\;
    remove $CCL$ segments on edges in $E^0$\ with minimum cost of segment $\geq curmax$\;
    update clients covered by $PSP$s in $P$ (to include $c_m$)\;
    $Q_1 \gets \{ p : p \in P \wedge p$ covers $c_m \}$\;
    $Q_2$ $\gets$ set of $PSP$s generated by $CCL(c_m)$ in $E^0$\;
    $P \gets P \cup Q_2$\;
    /* next consider $k$-candidates involving points in $Q_1 \cup Q_2$ */\;
    \While{$\exists$ new or updated $k$-candidate $Y$ not explored in $P$}{
        \If{$Y$ covers $c_1, ..., c_m$}{
        \If{ $Y.maxcost$ $< curmax$}{
                $kSP \gets Y$\;
                $curmax \gets$ $Y.maxcost$\;
        		\If{ $curmax$ $\leq Cost_S(c_{m+1})$}{
               	break\;}
       	}
      }
    }
    \If{$curmax > Cost_S(c_{m+1})$}{
         Return $curmax$, $kSP$}
}
}
}
\caption{$MinMax(G,C,S,k)$}
\label{alg:minmax}
\end{algorithm} 
%
%
The pseudocode is shown in Algorithm \ref{alg:minmax}.
Let the resulting minmax value be $minmax(G,C,S,k)$.
Let $kSP$ be the set of $k$ $PSP$s
returned. The positions of $kSP$, $\{ p.pos | p \in kSP\}$, are the locations for the $k$ new servers.
We examine combinations of $k$ $PSP$ points as candidate solutions. We call each such combination a
\textbf{$k$-candidate}.
Given a $k$-candidate $Y$, denote the highest cost among the $PSP$s in $Y$ by $Y.maxcost$, i.e.,
$Y.maxcost = max_{p \in Y} p.cost$.
According to our solution framework in Section \ref{sec:framework},
we try to incrementally cover a set of clients $\{c_1, ..., c_i \}$, from
$i = 1, ..., |C|$. Hence, we consider the clients $c$ with decreasing $Cost_S(c)$.
The key idea is to stop at some value of $i$ smaller than $|C|$.
We make use of the following lemma.


\begin{lemma}[early termination]
Let $C' = \{c_1, ..., c_m\}$.
If
$Cost_S(c_{m+1}) \leq \texttt{minmax}(G, C', S, k) < Cost_S(c_m)$,
then \\
$\texttt{minmax}(G,C,S,k) = \texttt{minmax}(G,C',S,k)$.
\end{lemma}



We begin the search with $\{c_1, ..,c_k\}$, since $k$ servers can reduce the costs of $k$ clients $\{c_1, ..,c_k\}$ to zero by choosing the client locations for placing the servers.
The for loop at Lines 7-25 iterates for clients from $c_{k+1}$ to $c_n$. The iteration for $c_m$ is incremental on
the previous iterations on
$c_1, ..., c_{m-1}$ avoiding repeated computation for the $PSP$s and $k$-candidates.
Each iteration adds a new $CCL(c_m)$ due to the consideration of a client $c_m$ (Line 9).
Lines 10-11 apply our pruning strategies, which will be explained shortly.
At Line 13, we store in $Q_1$ the existing $PSP$s which have been updated at Line 12 to include $c_m$.
At Line 14, we generate new $PSP$s from $CCL(c_m)$ in $E^0$
according to Lemma \ref{lem:PSP2}, and store them in $Q_2$.
A set $P$ is used to maintain the $PSP$s that have been generated and not pruned thus far.
At Line 17, the while loop looks for new or updated $k$-candidates. New $k$-candidates are sets of $k$ points in $P$ with at least one point from $Q_1 \cup Q_2$.
By `updated' $k$-candidates, we refer to new or old candidates which may have a new set of clients that they cover; in particular, $c_m$ can be added as a covered client.




%
%
%


Over the course of the computation,
let $kSP$ be
the current best $k$-candidate, and let $kSP.maxcost = curmax$.
We can safely dismiss $PSP$s with costs higher than
$curmax$.
This provides for dynamic pruning of $PSP$s at Line 10 and implicitly at Line 14,
which is shown to be highly effective in our empirical studies.

Similarly, given a linear piece of $CCL_{l}(c)$, if no point $p=(x,y)$ on $CCL_{l}(c)$ has a cost $y <= curmax$, we can dismiss $CCL_{l}(c)$ (Line 11).

%

At Line 22, if the maximum cost of the $PSP$s in the $k$-candidate is less than $Cost_S(c_{m+1})$, it means that
$Cost_S(c_{m+1})$ will be the minmax cost if we do not try to reduce the cost of $c_{m+1}$. To reduce the minmax cost, the solution must include at least a new $PSP$ from $CCL(c_{m+1})$ or an existing $PSP$ which will be updated at Line 12 to cover $c_{m+1}$. Hence, we can exit the while loop at Line 22, and process $c_{m+1}$ next.

At Line 24, if $curmax > Cost_S(c_{m+1})$, it implies that we cannot improve the minmax cost by considering clients
$c_{m+1}, ...., c_n$. Thus, the algorithm stops.



\begin{example}
Let us take the road network in Figure \ref{figG1} as an illustration. Suppose $k$ = 2 and $E^0$ = {[$v_3,v_4$],[$v_1,v_2$],[$v_1,v_3$]}. Assume that the client weights are the  following: $w(c_1)$=1, $w(c_2)$=1, $w(c_3)$=8, $w(c_4)$=3, $w(c_5)$=1 and $w(c_6)$=1. Note that $c_1$, $c_2$, $c_3$, $c_4$, $c_5$ and $c_6$ have their nearest servers as $s_2, s_1, s_2, s_2, s_1$ and $s_2$, respectively. We know that $c_1.dist$ = 8, $c_2.dist$ = 8, $c_3.dist$ = 1, $c_4.dist$ = 2, $c_5.dist$ = 4, $c_6.dist$ = 4. Thus, $Cost_S(c_1)$ = $w(c_1) \times c_1.dist$ = 1 $\times$ 8 = 8. Similarly, we have $Cost_S(c_2)$ = 8, $Cost_S(c_3)$ = 8, $Cost_S(c_4)$ = 6, $Cost_S(c_5)$ = 4, $Cost_S(c_6)$ = 4. Then, we have $Cost_S(c_1) \geq Cost_S(c_2) \geq Cost_S(c_3) \geq Cost_S(c_4) \geq Cost_S(c_5) \geq Cost_S(c_6)$. The client ordering is $c_1, c_2, c_3, c_4, c_5$ and $c_6$.

Since $k=2$, we build $CCL(c_1)$ and $CCL(c_2)$ on [$v_3,v_4$], [$v_1,v_2$], and [$v_1,v_3$] (Line 3). Next, we get $PSP$s $p_0$, $p_1$, $p_3$ on [$v_3,v_4$], $p_5$ on [$v_1,v_2$], $p_{10}$, $p_{12}$ on [$v_1,v_3$] (see Figures \ref{figG1}(c) to (e)). Thus, $P$ = $\{p_0$, $p_1$, $p_3$, $p_5$, $p_{10}$, $p_{12}\}$ (Lines 5 and 6).

When $m$ = 3, we know that the third client is $c_3$ and thus $curmax$ = $Cost_S(c_3)$ = 8 (Line 8). We build $CCL(c_3)$ (Line 9). No $PSP$ or $CCL$ is removed (Lines 10 and 11). No $PSP$ is updated since no $PSP$ in $P$ covers $c_3$ (Line 12) and  we get $PSP$ $p_8$ generated by $CCL(c_3)$ (Line 14). In the while loop, 2-$candidate$ $Y$ must contain $p_8$ (Line 17). When $Y$ = $\{p_1, p_8\}$, it covers $c_1, c_2$ and $c_3$ (Line 18). The fourth client is $c_4$. Since $curmax$ is updated to $Y.maxcost$ = $p_1.cost$ = 5 $<$ 6 = $Cost_S(c_4)$ (Lines 21 and 22), we exit the loop.

When $m$ = 4, we build $CCL(c_4)$ (Line 9). $CCL(c_1)$ on [$v_1,v_3$] and $p_{12}$ are removed (Lines 10 and 11) since $p_{12}.cost$ = 6 $\geq$ 6 = $curmax$. No $PSP$ is updated (Line 12). We get $PSP$s $p_6$ and $p_7$ generated by $CCL(c_4)$ (Line 14) (see Figure \ref{figG1}(d)). Now $P$ = $\{p_0$, $p_1$, $p_3$, $p_5$, $p_6$, $p_7$, $p_8$, $p_{10}\}$ and $Y$ should contain at least one $PSP$ in {$p_6$, $p_7$}. Only when $Y$ = $\{p_1$, $p_7\}$, $Y$ covers $c_1, c_2, c_3$ and $c_4$. Since $curmax$ = $Cost_S(c_4)$ = $6 > 5$ = $p_1.cost$ = $Y.maxcost$ (Line 19), $curmax$ is updated to $Y.maxcost$ = 5 (Line 21). The next client is $c_5$ and $curmax$ = 5 $>$ 4 = $Cost_S(c_5)$. Thus, at Line 25, $curmax > Cost_S(c_{m+1})$ holds, 5 and $\{p_1$, $p_7\}$ are returned as the solution by the algorithm.\done
\end{example}

Suppose the iterative process of Algorithm \ref{alg:minmax}
stops when $m=\gamma$ at Line 7.
Computing the $NLC$s and $CCL$s for the clients $c_1, ..., c_\gamma$ takes $O(\gamma |V| \log |V|)$ time.
Let $\rho$ be the number of eligible edges in these $NLC$s.
Let $\alpha$ be the maximum number of
$PSP$s for an edge, $\alpha = O(\gamma^2)$.
The time to compute the coverage for each $PSP$ is $O(\gamma)$. The total time to compute coverage is $O(\rho \alpha \gamma)$.
%
%
The running time of Algorithm \ref{alg:minmax} is $O( \gamma |V| \log |V| + \rho\alpha\gamma + k(\rho \alpha ) ^{k})$.
This shows that MinMax is in XP
 \cite{Flum2006Springer} and is computable for small $k$ values.
For memory requirement, we need to store the clients covered by the $PSP$s, which requires $O(\rho\alpha\gamma)$ storage. $k$-candidates are computed on the fly.
The memory complexity is thus $O(\rho\alpha\gamma)$.

\section{Optimization}
\label{sec:enhancedMinMax}



In the previous subsection, we describe an algorithm solving the MinMax problem which takes $O( \gamma |V| \log |V| + \rho\alpha\gamma + k(\rho \alpha ) ^{k})$ time, where $O(\rho \alpha)$ is the number of $PSP$s processed, and $O((\rho \alpha)^{k})$ is the number of $k$-candidates considered, which is the dominating factor. In this subsection, we introduce enhancement techniques to optimize the algorithm by reducing the computation cost of $PSP$s and $k$-candidates.
The proposed strategies reduced the number of $PSP$s to
20 or less for $k \leq 10$ in our real datasets, which makes the optimal solution more  effective.

\subsection{Early Termination of Iterations}
\label{sec:strategy}

In the iteration for client $c_m$ in Algorithm \ref{alg:minmax}, we compute $PSP$s generated by $CCL(c_m)$ to include in $P$, update clients covered by $PSP$s and look for $k$-candidates in $P$. If we can jump to the next iteration for $c_{m+1}$ before these steps, we may reduce the computation cost substantially.
To achieve this, we introduce two strategies that allow us to go to next iteration before computing $PSP$s for the current iteration. The pseudocode that incorporates these strategies is given in Algorithm \ref{alg:minmax2}.


\subsubsection{\bf Strategy 1: $kSP$ based early termination}

Consider the iteration for $c_m$. After building $CCL(c_m)$ on eligible edges, we check whether any $PSP$ in the current best $k$-candidate $kSP$ can cover $c_m$ at Line 14 in Algorithm \ref{alg:minmax2}. If one of the $PSP$s in $kSP$ covers $c_m$ and $kSP.maxcost$ $<Cost_{S}(c_{m+1})$, we can keep $kSP$ and jump to the next iteration to process $c_{m+1}$. This is because we can be sure that the $minmax$ cost would be either $Cost_{S}(c_{m+1})$, or that a smaller cost can be found with a solution set that also covers client $c_{m+1}$.

If we jump to the next iteration early in the iteration of $c_{m}$, $PSP$s to be generated by $CCL(c_m)$ are not computed and $PSP$s in $P$ are not updated to include coverage of $c_m$. Thus, we use a set $N$ to keep track of clients whose iterations are terminated early. As we continue with the for loop and come to an iteration where we cannot jump early, say in the iteration of $c_{i}$ where $i>m$, we compute and update $PSP$s for clients in set $N$ at Lines 23 and 24. The value $curmax$ typically becomes much smaller than $Cost_{S}(c)$ for $c \in N$ if we jump multiple iterations after $c$. Many $CCL$s can be removed and many $PSP$s can be disregarded. The number of $PSP$s to be processed can thus be significantly reduced.

\subsubsection{\bf Strategy 2: Virtual $PSP$s based early termination}

In iteration $m$, for each edge $\ell$, we create a \textbf{virtual} $PSP$ $p_\ell$ with $p_\ell.cost = Cost_{S}(c_{m+1})$. $p_\ell$ can be considered as a point $p = (x,y)$ with $x$ undefined and $y = Cost_{S}(c_{m+1})$. We say that $p_\ell$ is a $PSP$ of edge $\ell$.
Define the set of clients covered by $p_\ell$ to be the set of all clients $c$ such that the maximum cost of points on $CCL_{\ell}(c)$ is smaller than $Cost_{S}(c_{m+1})$.
We call the normal (non-virtual) $PSP$s the actual $PSP$s.

\begin{figure}[h!]
\begin{center}
\hspace*{-3mm}
\includegraphics[height=0.9in]{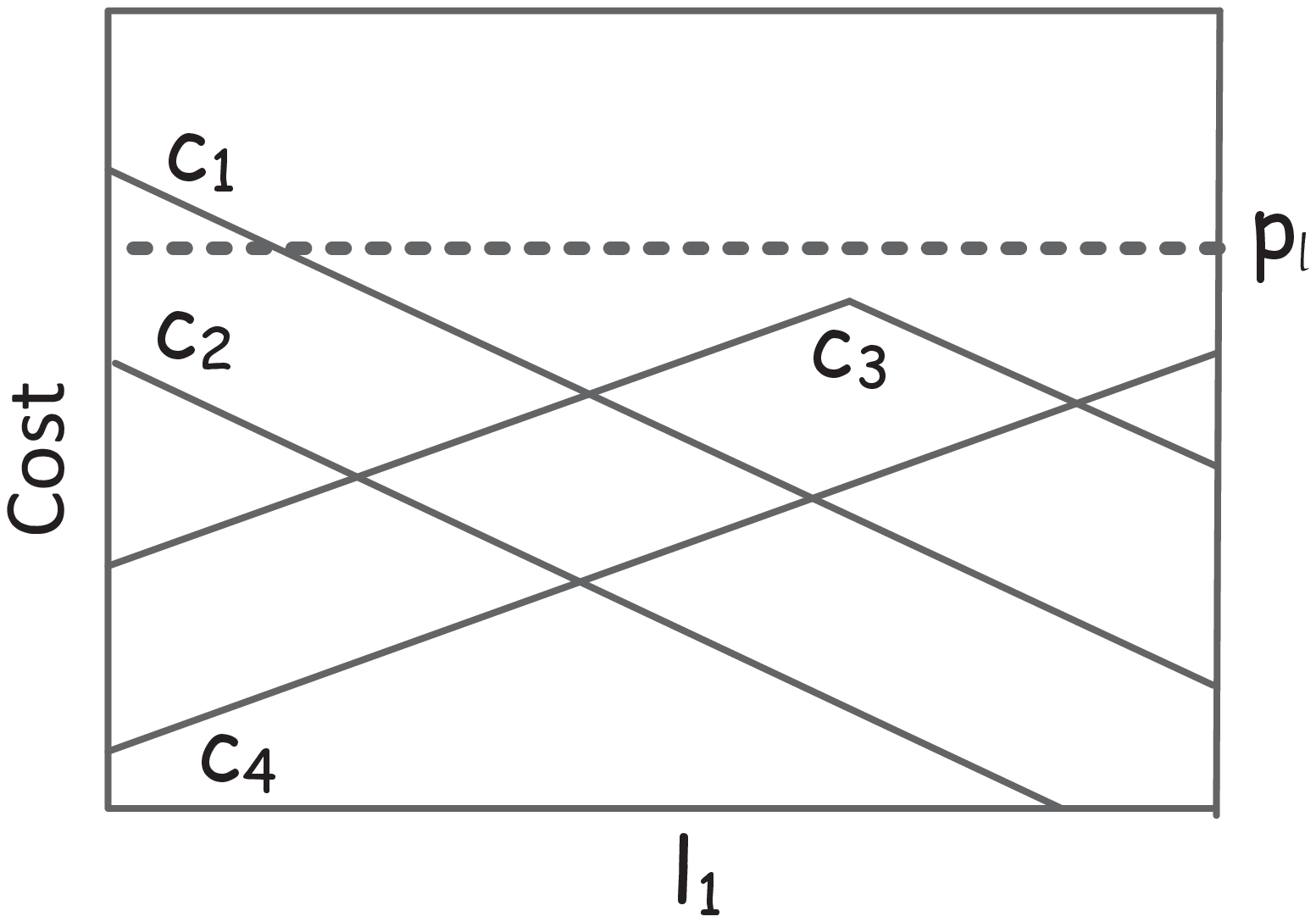}
\includegraphics[height=0.9in]{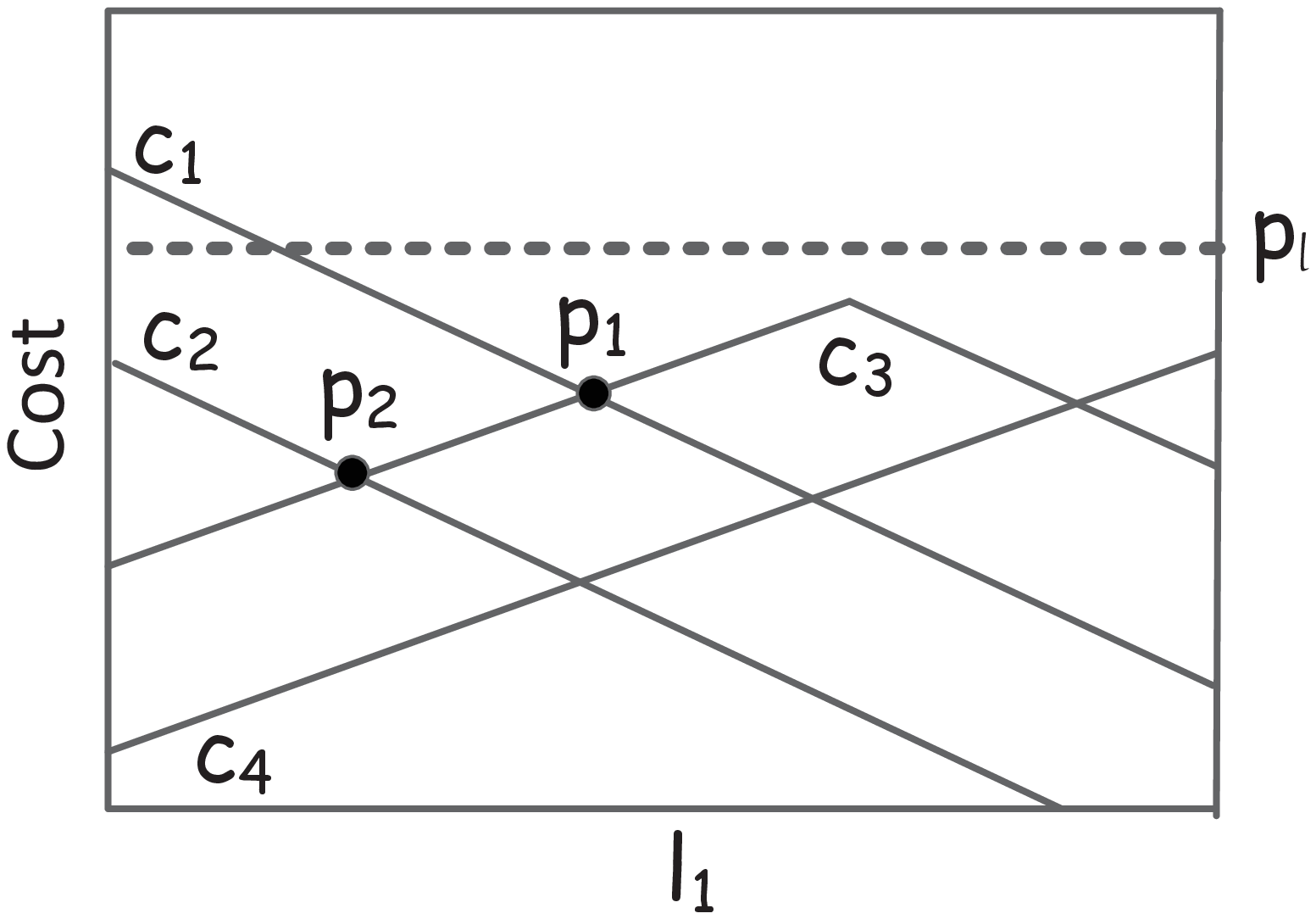}
\\
(a) \hspace*{1.3in} (b)

\includegraphics[height=0.8in]{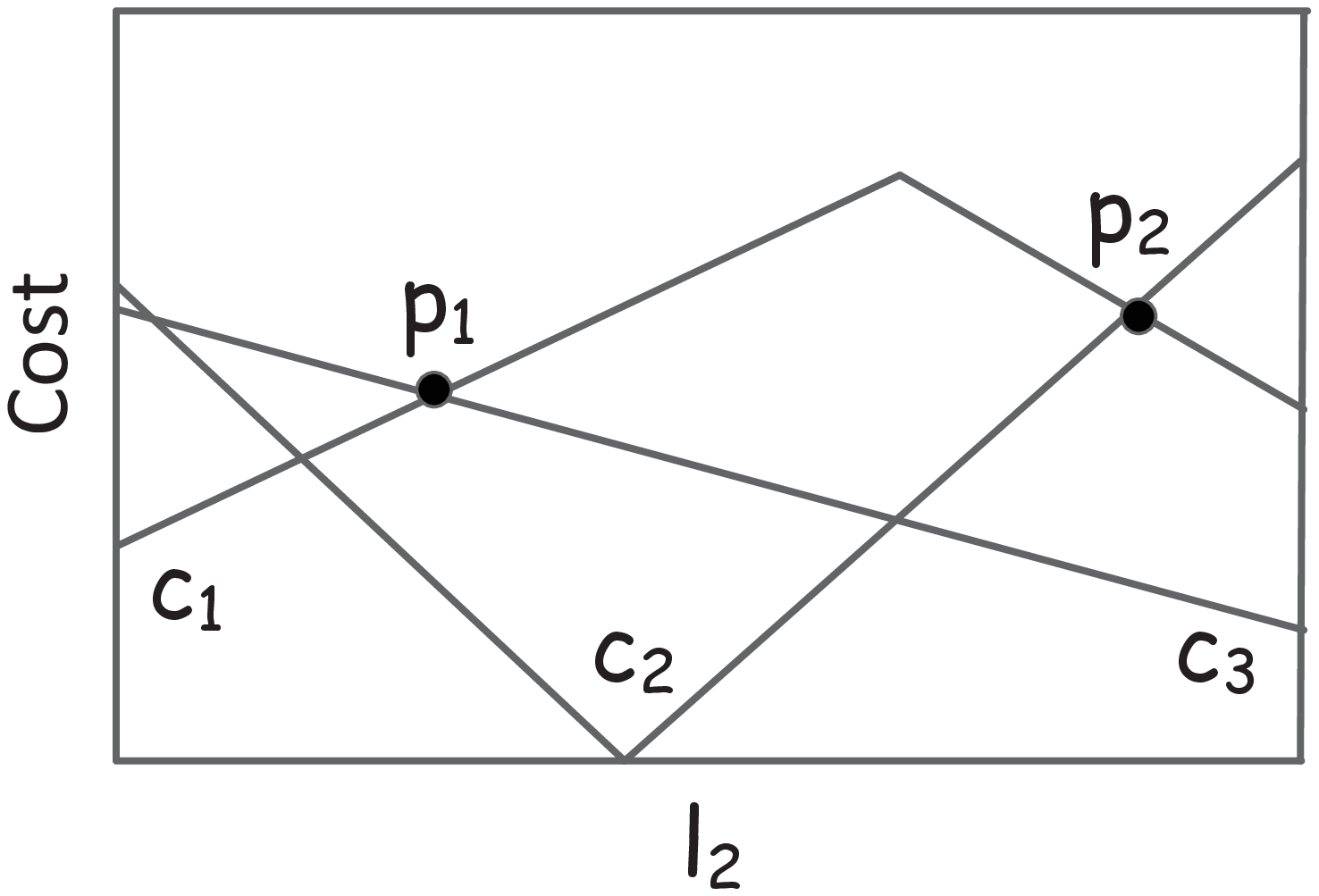}
\includegraphics[height=0.8in]{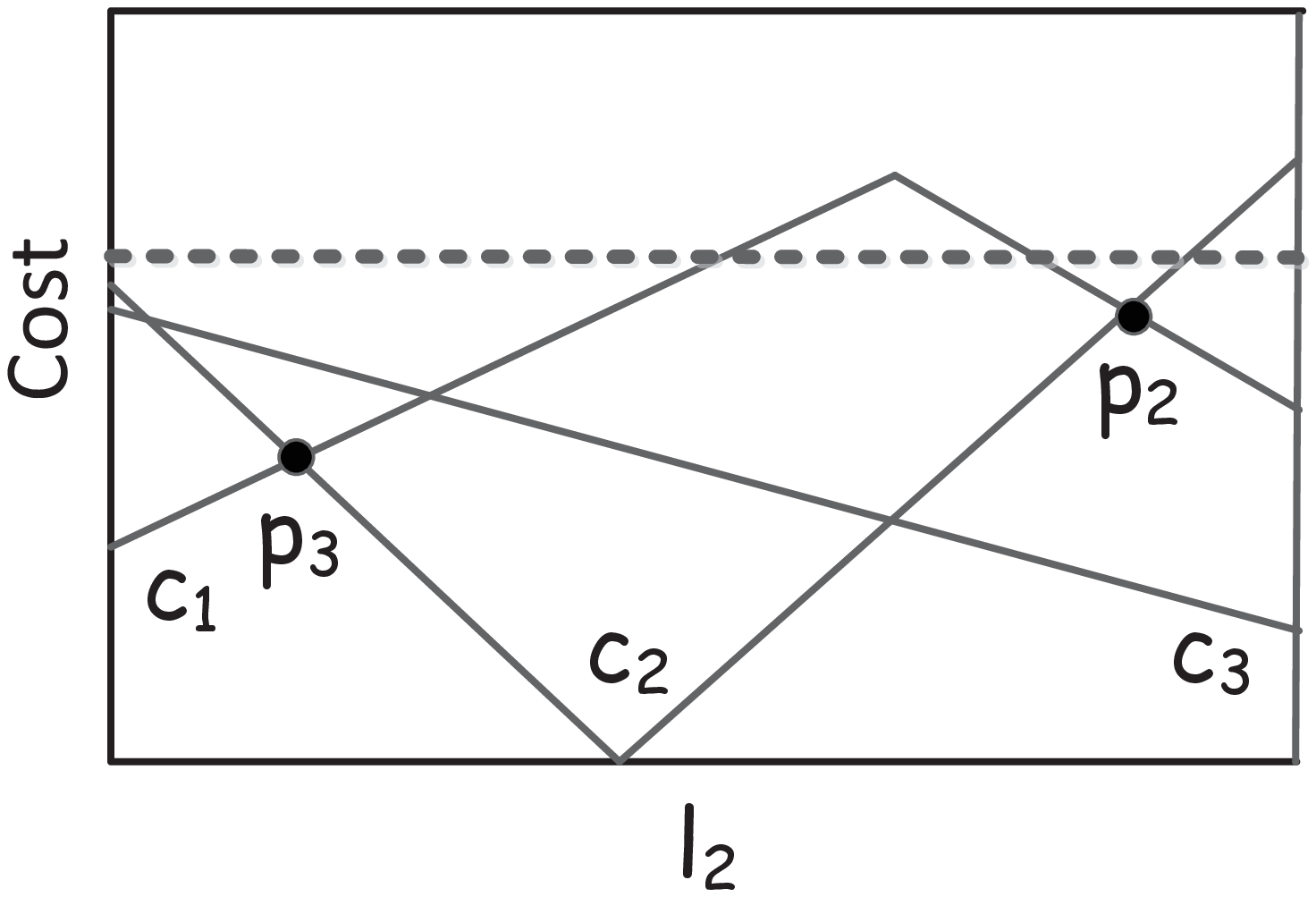}
\\
(c) \hspace*{1.3in} (d)

\end{center}
\vspace*{-3mm}
\caption{Illustrations for the enhancement strategies.
(a) virtual $PSP$ $p_l$ for edge $l_l$ covers $c_2, c_3$. (b) $p_2$ is pruned by $P_{edge}$. (c) $p_1$ and $p_2$ both cover $c_1, c_2, c_3$; $p_2$ is pruned. (d) $p_3$ is pruned when compared with $p_2$. }
\vspace{-2mm}
\label{eg:fig4}
\end{figure}

In Figure \ref{eg:fig4}(a), the dotted line shows $p_{l}$ which is the virtual $PSP$ of edge $l_1$. The cost of $p_{l}$ equals $Cost_{S}(c_{m+1})$. $p_{l}$ covers $c_2$, $c_3$, and $c_4$. There exist some points on $CCL(c_1)$ with costs greater than $p_{l}.cost$, so $p_{l}$ does not cover $c_1$.

We build a $PSP$ set $P_{edge}$ which consists of all virtual $PSP$s of eligible edges at Lines 7 and 16 in Algorithm \ref{alg:minmax2} and select $k$-candidates in $P_{edge}$ at Line 17. Each selection is a combination of $k$ virtual $PSP$s in $P_{edge}$. The cost of each $PSP$ in $P_{edge}$ is $Cost_{S}(c_{m+1})$.

\begin{lemma}
In the iteration of $c_m$, given a virtual $k$-candidate in $P_{edge}$ that covers $c_1$,...,$c_m$, there must exist a
$k$-candidate $Z$ with $k$
actual $PSP$s that cover $c_1$,...,$c_m$ with
$Z.maxcost \leq Cost_{S}(c_{m+1})$.
\label{lem:minmax2}
\end{lemma}

\vspace*{-3mm}

{\small PROOF}:
Given a virtual $k$-candidate $Z_v$ in $P_{edge}$ which covers the client set $C_m = \{c_1, ..., c_m\}$. For each $p_\ell \in Z_v$ on edge $\ell$, let $Z_\ell$ be the set of clients covered by $p_\ell$ where $Z_\ell \subseteq C_m$.
Consider $c_i \in Z_\ell$, by the definition of $p_\ell$, and since $Cost_S(c_i) > Cost_S(c_{m+1})$ by the sorted order of all clients,
it follows that $CCL_\ell(c_i)$ must be a single linear segment with two end points.
Thus, the point $h_\ell$ with the highest cost among all $CCL_\ell(c_i)$ for $c_i \in Z_\ell$ must cover all clients in $Z_\ell$. The set of such points $h_\ell$ forms a
$k$-candidate $Z$ that covers $c_1$, ..,$c_m$ with
$Z.maxcost \leq Cost_S(c_{m+1})$. The lemma follows from Lemma \ref{lem:PSP2} since an optimal $k$-candidate can be obtained from the actual $PSP$s.
\done

\medskip

Based on Lemma \ref{lem:minmax2}, at Line 18, if we get a $k$-candidate from $P_{edge}$ that covers $c_1$,...,$c_m$, we can update $curmax$ to $Cost(c_{m+1})$ and jump to the next iteration without computing $PSP$s for $c_m$. The size of $P_{edge}$ is $|E^{0}|$, which is much smaller than the size of $P$, hence the processing is much faster.

In our empirical study, we can find a $k$-candidate in $P_{edge}$ that meets the requirements in most cases.
The reason is that in the iteration of $c_m$, only clients $c$ with $CCL_{l}(c)$s containing some point $p$ with $p.cost \geq Cost(c_{m+1})$ are not covered by $p_\ell$. Since $CCL$s with a minimum cost greater than $curmax$ are removed at Line 12, and $curmax \leq Cost_S(c_m)$, we only consider $CCL$s containing some point $p$ such that $p.cost \leq curmax \leq Cost_{S}(c_m)$.
As the clients are sorted, $Cost_S(c_{m+1})$ is typically close to $Cost_S(c_m)$, thus, $p.cost$ is also likely to be $\leq Cost(c_{m+1})$.
It is therefore likely for $p_\ell$ to cover all clients $c$ such that $CCL_{l}(c)$ is on edge $l$.
As with Strategy 1, we keep $c_m$ in set $N$ if we skip computations of $PSP$s due to $c_m$, to be handled in a future iteration.

\subsection{PSPs Pruning}
\label{sec:PSPpruning1}

The complexity in computing $k$-candidates depends on the size of $P_{edge}$ at Line 17 and the size of $P_{edge} \cup P$ at Line 27. In this subsection, we introduce two strategies for reducing the number of $PSP$s, which can greatly speed up the algorithm.


\subsubsection{\bf Strategy 3: Pruning by PSP comparison}

We identify 2 properties of $PSP$s to eliminate redundant $PSP$s from the computed $PSP$s.


\begin{property}
If there exist two $PSP$s that cover the same set of clients, we can safely disregard the one with a higher cost when computing $k$-candidates.
\end{property}


{\small RATIONALE FOR PROPERTY 1}:
Consider the example in Figure \ref{eg:fig4}(c). $p_1$ and $p_2$ are two $PSP$s on edge $l_2$, and both cover $c_1$, $c_2$ and $c_3$. $p_1.cost < p_2.cost$. Suppose we get a $k$-candidate $Z$ that contains $PSP$ $p_2$, we can replace $p_2$ by $p_1$ directly. If $p_2.cost = Z.maxcost$, then we can replace $p_2$ by $p_1$ to convert $Z$ to $Z'$, and $Z'.maxcost$ will be $p_1.cost$, which is less than $p_2.cost$. If $p_2.cost \neq Z.maxcost$, then $Z'.maxcost$ will be equal to $Z.maxcost$ after replacing $p_2$ by $p_1$.
\done


\begin{property}
Suppose $p_1$ and $p_2$ are two $PSP$s on the set of eligible edges, the set of clients covered by $p_1$ is $C_1$ and that covered by $p_2$ is $C_2$. In the iteration of $c_m$, if $p_1.cost< p_2.cost < Cost_{S}(c_{m+1})$ and $C_1 \subset C_2$, we can disregard $p_1$ when computing $k$-candidates.
\end{property}


{\small RATIONALE FOR PROPERTY 2}:
Consider Figure \ref{eg:fig4}(d). $p_2$ and $p_3$ are two $PSP$s on edge $l_2$. $p_2$ covers $C_2$ = $\{ c_1, c_2, c_3 \}$. $p_3$ covers $C_3$ = $\{ c_1, c_2 \}$. Suppose the dashed line shows $Cost_{S}(c_{m+1})$, the cost of the client in the next iteration. $C_3 \subset C_2$ and  $p_3.cost< p_2.cost < Cost_{S}(c_{m+1})$. Consider a $k$-candidate $Z$ that contains $PSP$ $p_3$, we can replace $p_3$ by $p_2$, and get $Z' = Z \setminus \{p_3\} \cup \{p_2\}$. Hence, clients covered by $Z$ are also covered by $Z'$. If $Z.maxcost$ $>$
$Cost_{S}(c_{m+1})$, then $Z'.maxcost=Z.maxcost$. Hence, if $Z$ is an optimal solution, $Z'$ is also an optimal solution.
If $Z.maxcost$ $\leq$ $Cost_{S}(c_{m+1})$, then $Z'.maxcost$ is still less than or equal to $Cost_{S}(c_{m+1})$ since $p_3.cost< p_2.cost < Cost_{S}(c_{m+1})$. Thus, neither $Z$ nor $Z'$ will be returned as $kSP$ at
Lines 34 and 35. Therefore, we can disregard $p_3$ when computing $k$-candidates.
Note that while $p_3$ is disregarded in this iteration, it can be used for computing $k$-candidate in a future iteration for some $c_i$, $i > m$, if $p_3$ is updated at Line 23 to cover $c_i$ while $p_2$ cannot cover $c_i$.
\done

Note that although the example above is based on only one edge, in general
we apply these 2 properties to $PSP$s in $P_{edge}$ at Line 17 and $P_{edge} \cup P$ at Line 27 on all eligible edges. That is, the two $PSP$s involved can be on two different edges.


\subsubsection{ \bf Strategy 4: $P_{edge}$ based pruning}

When we cannot jump to the next iteration by Strategy 1 or 2, we will compute $PSP$s generated by $CCL(c)$ for $c \in N$ at Line 24. With the set $P_{edge}$, we can reduce the number of $PSP$s to be computed in the following way.
For each edge $\ell$,
a $CCL_{\ell}(c)$ is said to be \textbf{active} if $CCL_{\ell}(c)$ contains a point $p$ with $p.cost > Cost_{S}(c_{m+1})$ in the iteration of $c_{i}$. Otherwise, $CCL_{\ell}(c)$ is \textbf{inactive}. While computing $PSP$s at Line 24, we check whether a $PSP$ $p$ is generated by intersecting inactive $CCL_{\ell}(c)$s, and if so, $p$ is dismissed. Clients covered by such a $PSP$ must be also covered by $p_\ell$, the virtual $PSP$ of edge $\ell$. Since $p_\ell$ will be considered at Line 27, based on a similar rationale as that for Property 2, $p$ can be dismissed.

Consider Figure \ref{eg:fig4}(b). $p_{l}$ is the virtual $PSP$ on the edge $l_1$, and it covers $C_{l_1} = \{c_2, c_3, c_4\}$. $CCL_{l_1}(c_1)$ is active. $CCL_{l_1}(c_2)$ and $CCL_{l_1}(c_3)$ are inactive. $p_1$ is an intersection of $CCL_{l_1}(c_1)$ and $CCL_{l_1}(c_3)$, which covers $C_1 = \{c_1, c_2, c_3 ,c_4\}$. $p_2$ is an intersection of $CCL_{l_1}(c_2)$ and $CCL_{l_1}(c_3)$, which covers $C_2$ = $\{ c_2, c_3, c_4 \}$. $C_1 \not\subseteq C_{l_1}$ and $C_2 \subseteq C_{l_1}$. We can thus dismiss $p_2$.



\begin{algorithm}[t]
\SetKwInOut{input}{Input}\SetKwInOut{output}{Output}
{\scriptsize
\input{$G$,$C$,$S$,$k$, eligible edges $E^0$, sorted $c_1, ..., c_n$}
\output{minmax cost: $\texttt{minmax}(G,C,S,k), kSP$}

\Begin{
	$P \gets \emptyset; P_{edge} \gets \emptyset$;
	$N \gets \emptyset$\;
	
	\For{$m = 1, ..., k$}{
 	    build $CCL(c_m)$ on each edge in $E^0$\;
	}
	
	Compute set $Q$ of $PSP$s and clients covered by each $PSP$ on $E^0$\;
	$P \gets P \cup Q$\;
	For each $\ell$ in $E^0$, compute clients covered by $p_\ell$ and $P_{edge} \gets \{p_\ell\}$\;
	\For{$m = k+1, ..., n$}{
		$curmax \gets Cost_S(c_{m})$ \;
		$N \gets N \cup \{c_m\}$\;
   		build $CCL(c_m)$ on edges in $E^0$\;
    	remove $PSP$s in $P$ with cost $\geq curmax$\;
    	remove $CCL_{\ell}(c)$ where $c\in C$ on each edge $\ell$  in $E^0$\ with a minimum cost $\geq curmax$\;
		\If{ $kSP.maxcost$ $\leq Cost_{S}(c_{m+1})$ and $kSP$ covers $c_m$ }{
		 	$continue$    \Comment{/* Strategy 1 */}
		}
    	
   		update $p_\ell.cost$ and clients covered by $p_\ell$ for each $p_\ell$ in $P_{edge}$\;
		\While{$\exists$ updated $k$-candidate $Z$ not explored in $P_{edge}$}{
        	\If{$Z$ covers $c_1, ..., c_m$}{
                	$curmax \gets$ $Cost_S(c_{m+1});$ $kSP \gets$ $Z$\;
                break\;
      		}
    	}
    	\If{$curmax = Cost_S(c_{m+1})$}{
		 	$continue$       \Comment{/* Strategy 2 */}
		}

		update clients covered by $PSP$s in $P$ (to include clients in $N$)\;
    	$Q$ $\gets$ set of $PSP$s generated by $CCL(c), c \in N$\;
    	$N \gets \emptyset $\;
    	$P \gets P \cup Q$\;
    	
		\While{$\exists$ new or updated $k$-candidate $Z$ not explored in $P_{edge} \cup P$}{
        	\If{$Z$ covers $c_1, ..., c_m$}{
        		\If{ $Z.maxcost$ $< curmax$}{
                	$kSP \gets$ $Z$\;
                	$curmax \gets$  $Z.maxcost$\;
                	\If{ $curmax$ $\leq Cost_S(c_{m+1})$}{
            			break\;
           			}
            	}
      		}
    	}
    	\If{$curmax > Cost_S(c_{m+1})$}{
         	Return $curmax, kSP$
    	}	
   	}

}
}
\caption{$QuickMinMax(G,C,S,k)$}
\label{alg:minmax2}
\end{algorithm}

\subsection{Algorithm \ref{alg:minmax2}}

We call the enhanced MinMax algorithm QuickMinMax. It is shown
in Algorithm \ref{alg:minmax2}. 
Strategy 1 is carried out at Lines 14-15.
Since $|P_{edge}|$ is much less than $|P|$, we first select $k$-candidates from $P_{edge}$ at Line 17, and execute Strategy 2 at Lines 17-22.
If either Strategy 1 or 2 succeeds, note that $c_m$ is added to $N$ at
Line 10,
and we jump to the next iteration to process $c_{i+1}$.
Otherwise we cannot terminate early, so $N$ is processed at Lines 23 and 24 and reset to empty at Line 25.
The remaining processing is similar to Algorithm \ref{alg:minmax}, except that Strategy 3 and Strategy 4 are activated at Lines 17 and 27, and at Line 24, respectively.


\section{Empirical Studies}

\label{sec:exp}

In this section, we evaluate the performance of our proposed algorithms.
We run all experiments on a machine with a 3.4Ghz Intel Core i7-4770 CPU and 16 GB RAM, running Ubuntu 12.04 LTS Linux OS. All algorithms were implemented in C++ and compiled with GNU c++ compiler.

We use three real world road network datasets: SF, NYC and FLA,
for San Francisco, New York City, and Florida, respectively.
SF contains
174,955 vertices and 223,000 edges.
NYC contains
264,346 vertices and 733,846 edges and
FLA contains 1,070,376 vertices and 2,712,798 edges.
In our experiment, clients and servers are generated randomly on all edges.
Each client is associated with a weight, generated randomly from a Zipf distribution with a skewness parameter of
$\alpha > 1$.
The default setting is as follows:
For SF:
$|S|$= 200,
$|C|$ = 100,000.
For NYC:
$|S|$ = 500,
$|C|$ = 300,000.
For FLA:
$|S|$ = 1000,
$|C|$ = 600,000.
$\alpha = \infty$, meaning a unit weight for each client.
The default setting for $|E^0|/|E|$ is 10\%.

We measure the quality and runtime by varying different parameters, including the number of new servers $k$,
the number of clients $|C|$, number of existing servers $|S|$,
and
the Zipf factor $\alpha$ in the skewness of client weights.


%
%


\subsection{Comparison of the Algorithms}

We analyze the effects of different parameters on the MinMax algorithms.
We compare results of our approximation algorithm (\textbf{Approx})
(Algorithm \ref{alg:CodeApprox2}) and optimal algorithm (\textbf{Opt})
(Algorithm \ref{alg:minmax2}) with the results from the best-known greedy algorithms (\textbf{Greedy}) in \cite{chen2014efficient}.
Note that the results of \cite{chen2014efficient} and \cite{xiao2011optimal}
are identical since they
compute the same optimal solution for a single new server, and
the corresponding approximation algorithms repeatedly select a single new server until $k$ new server locations are chosen.
However, the algorithm in \cite{chen2014efficient}
improves on the computation time compared to \cite{xiao2011optimal},
hence we only report the runtime of the greedy algorithms in \cite{chen2014efficient}.

Let the minmax cost prior to adding any new server be $Max = max(Cost_S(c)|c \in C)$,
$P$ be an optimal or approximate MinMax solution with $k$ locations.
Let $kMax = max(Cost_{S \cup P}(c)|c \in C)$.
Define $Gain = Max - kMax$, and
Gain Ratio $GR = Gain/Max$.
To quantify the advantage of the optimal solution over the approximation algorithm, we measure
the relative error of Approx as $Err = (O - A)/O$,
where $A$($O$) is the $Gain$ from Approx
(Opt).

\renewcommand{\tabcolsep}{5pt}
\begin{table}[h!]
\begin{scriptsize}
\begin{tabular}{|l||r|r|r|r|r|r|}
  \hline
k & 1 & 2  & 4 & 6 & 8  & 10 \\
  \hline
  \hline
Gain (Opt) & 37.3 & 89.0  & 1277.8 & 1722.4 & 1800.5 &1986.5 \\
Gain (Approx) & 37.3 & 89.0  & 1041.3 & 1277.8 & 1499.2 & 1546.8 \\
Gain (Greedy) & 37.3 & 66.3  & 89.0 & 124.0 & 192.1 & 221.0 \\
\hline
GR (Opt) (\%) & 1.1 & 2.6  & 37.4 & 50.4 & 52.7 & 58.1 \\
GR (Approx) (\%) & 1.1 & 2.6  & 30.5 & 37.4 & 43.9 &45.3 \\
GR (Greedy) (\%) & 1.1 & 1.9  & 2.6 & 3.6 & 5.6 &6.5 \\
\hline
Relative Error (Approx) & 0.00 & 0.00  & 0.18 & 0.26 & 0.16	 & 0.22\\
Relative Error (Greedy) & 0.00 & 0.25  & 0.93 & 0.92 & 0.89 & 0.89\\
  \hline
Time (Opt)(s) & 0.8 & 1.9 & 3.1 & 28.4 & 132.9 & 317.0\\
Time (Approx)(s) & 0.3 & 0.3  & 0.4 & 0.5 & 0.5 & 0.6 \\
Time (Greedy)(s) & 0.6 & 0.7  & 0.7 & 0.8 & 0.9 & 0.9 \\
\hline
\end{tabular}
\end{scriptsize}
\vspace*{-2mm}
\caption{Gain, Gain Ratio (GR), Relative Error, and runtime for MinMax (Opt:Optimal solution; Approx: approximate solution; Greedy: greedy solution) with unit weights on SF}
\label{tab:quality1}
\end{table}
\vspace*{-2mm}

\begin{table}[h!]
\begin{scriptsize}
\begin{tabular}{|l||r|r|r|r|r|r|}
  \hline
k & 1 & 2 & 4 & 6 & 8 &10  \\
  \hline
  \hline
Gain (Opt) & 5.9 & \hspace{0.05cm}40.1  & \hspace{0.05cm}77.1 & \hspace{0.05cm}217.2 & \hspace{0.05cm}313.3 &\hspace{0.05cm} 324.1\\
Gain (Approx) & 5.9 & 40.1 & 75.7 & 106.9 & 217.2 & 281.8\\
Gain (Greedy) & 5.9 & 6.0 & 28.8 & 41.0 & 59.1 & 68.8\\
\hline
GR (Opt) (\%) & 0.4 & 2.8  & 5.2 & 15.0 & 21.6 & 22.4\\
GR (Approx) (\%) & 0.4 & 2.8  & 5.3 & 7.4 & 15.0 & 19.4\\
GR (Greedy) (\%) & 0.4 & 0.4  & 2.0 & 2.8 & 4.1 & 4.7\\
\hline
Relative Error (Approx) & 0.00 & 0.00  & 0.02 & 0.50 & 0.30 & 0.13\\
Relative Error (Greedy) & 0.00 & 0.85  & 0.63 & 0.81 & 0.81 &0.78 \\
  \hline
Time (Opt)(s) & 1.4 & 1.6 &  2.04 & 22.2 & 301.9 &401.8\\
Time (Approx)(s) & 0.5 & 0.6  & 0.9 & 1.1 & 1.3 & 1.5 \\
Time (Greedy)(s) & 1.2 & 1.3 & 1.5 & 1.6 & 1.8 & 2.0\\
\hline
\end{tabular}
\end{scriptsize}
\vspace*{-2mm}
\caption{Gain, Gain Ratio (GR), Relative Error, and runtime for MinMax
with unit weights on NYC}
\label{tab:quality2}
\end{table}
\vspace*{-2mm}

\begin{table}[h!]
\begin{scriptsize}
\begin{tabular}{|l||r|r|r|r|r|r|}
  \hline
k & 1 & 2 & 4 & 6 & 8 & 10 \\
  \hline
  \hline
Gain (Opt) & \hspace{0.05cm}45.5 & \hspace{0.05cm}241.3 &  \hspace{0.05cm}249.8 & \hspace{0.05cm}275.1 & \hspace{0.05cm}286.7 & \hspace{0.05cm}296.8 \\
Gain (Approx) & 45.5 & 82.4 &  248.6 & 272.5 & 279.6 & 285.2\\
Gain (Greedy) & 45.5 & 54.4 &  56.7 & 61.1 & 62.4 & 64.3\\
\hline
GR (Opt) (\%) & 6.1 & 32.1 & 33.3 & 36.6 & 38.2 &39.5 \\
GR (Approx) (\%) & 6.1 & 11.0 & 33.1 & 36.3 & 37.2 & 38.0 \\
GR (Greedy) (\%) & 6.1 & 7.2 & 7.6 & 8.1 & 8.3 &8.5 \\
\hline
Relative Error (Approx) & 0.00 & 0.66  & 0.01 & 0.01 & 0.02 & 0.04\\
Relative Error (Greedy) & 0.00 & 0.77  & 0.78 & 0.78 & 0.78  & 0.78 \\
  \hline
Time (Opt)(s) & 5.4 & 28.9 &  32.6 & 42.2 & 70.3 & 190.9\\
Time (Approx)(s) & 1.9 & 2.2 & 2.6 & 3.1 & 3.5 & 3.9 \\
Time (Greedy)(s) & 4.6 & 4.7 & 5.0 & 5.4 & 5.7 & 6.2\\
\hline
\end{tabular}
\end{scriptsize}
\vspace*{-4mm}
\caption{Gain, Gain Ratio (GR), Relative Error, and runtime for MinMax
with unit weights on FLA}
\vspace*{-2mm}
\label{tab:quality3}
\end{table}

For comparison of the optimal solution and approximate solution with the greedy solution, we show the $Gain$ and Gain Ratios $GR$ in Tables
\ref{tab:quality1}, \ref{tab:quality2}, \ref{tab:quality3} and \ref{tab:quality4}.
The results clearly show the advantages of the optimal solution and approximate solution compared to the greedy algorithm.
For example, to achieve the same level of gain as
2 new servers in the optimal solution,
4, 6 and over 10 servers will be needed from the greedy solution in SF, NYC, and FLA, respectively.
With our target applications, it would incur a large and unnecessary cost for the user to build more than 10 servers instead of 2, thus the longer running time needed for the optimal solution is well justified for the saving in this expense. The $Gain$ and $GR$ of $Approx$ are always better than $Greedy$ and the relative error is small. $Approx$ has a poor gain with FLA for $k=2$ because in FLA the clients are more dispersed and it is harder to achieve 
near optimal result when the locations are limited to the client sites.
Note that for clients with skewed weights, $Opt$ takes about the same time as $Greedy$ and $Approx$ as shown in Table \ref{tab:quality4}.

\renewcommand{\tabcolsep}{3pt}
\begin{table}[h!]
\begin{scriptsize}
\begin{tabular}{|l||r|r|r|r|r|r|}
  \hline
k & 1 & 2 & 4 & 6 & 8 & 10   \\
  \hline
  \hline
Gain (Opt) & 974.0 & 3786.0  & 5845.9 & 8841.8 & 9272.9 & 10268.4 \\
Gain (Approx) & 974.0 &  3786.0  & 5845.9 & 7454.2 & 8222.8 & 8701.9\\
Gain (Greedy) & 974.0 &  1691.0  & 3786.0 & 5344.5 & 5845.9 & 6094.3\\
\hline
GR (Opt) (\%) & 6.0 & 23.5 &  36.2 & 54.8 & 57.5 & 63.7 \\
GR (Approx) (\%) & 6.0 & 23.5 & 36.2 & 46.2 & 51.0 & 54.0\\
GR (Greedy) (\%) & 6.0 & 10.5 & 23.5 & 33.1 & 36.2 & 37.8\\
\hline
Relative Error(Approx)& 0.00 & 0.00  & 0.00 & 0.16 & 0.11 & 0.15\\
Relative Error(Greedy)& 0.00 & 0.55 & 0.35 & 0.40 & 0.37 & 0.41 \\
\hline
Time (Opt)(s) & 0.8 & 0.8  & 0.9 & 1.8 & 2.3 & 4.0\\
Time (Approx)(s) & 0.3 & 0.3 & 0.4 & 0.5 & 0.5 & 0.6 \\
Time (Greedy)(s) & 0.7 & 0.7 & 0.8 & 0.8 & 0.9 & 0.9\\
  \hline
\end{tabular}
\end{scriptsize}
\vspace*{-2mm}
\caption{Gain, Gain Ratio (GR), Relative Error, and runtime for MinMax
with skewed weights ($\alpha = 2$) on SF}
\label{tab:quality4}
\end{table}

We analyze the effect of $k$ and the Zipf parameter $\alpha$ on the $Gain$ in Table \ref{tab:quality5}. The optimal solution consistently generated better solutions compared to the approximation method. Since the greedy algorithm always takes more time and returns worse result compared to the approximation algorithm, we only show the $Gain$ of $Opt$ and $Approx$.

\renewcommand{\tabcolsep}{5pt}
\begin{table}[h!]
\begin{center}
\begin{scriptsize}
\begin{tabular}{|l|r||r|r|r|r|r|r|}
\hline
  \multicolumn{2}{|l||}{ \ } &
  \multicolumn{3}{c|}{Opt Gain (SF)}\\
  \hline
$\alpha$ & Max & $k=2$  & $k=4$  & $k=8$   \\
  \hline
  \hline
2 & 16129 & 3786.0  &  5845.9  &  9272.9   \\
4  & 12124 & 5647.4 & 6578.1 &  7597.8   \\
6 & 6451 & 2133.8 & 2754.3 & 4018.4 \\
$\infty$ & 3417 & 89.0 & 1277.8 & 1800.5\\
  \hline
\end{tabular}
\begin{tabular}{|r|r|r|r|}
  \hline
  \multicolumn{3}{|c|}{Approx Gain (SF)}\\
  \hline
 $k=2$  & $k=4$  & $k=8$   \\
  \hline
  \hline
 3786.0  &  5845.9  &  8222.8\\
 3449.1 & 5647.4 &  7187.0 \\
 1246.6 & 2133.8 & 3034.6 \\
 89.0 & 1041.3 & 1499.2 \\
  \hline
\end{tabular}
\\
{ \ \ }
\\
\begin{tabular}{|l|r||r|r|r|r|r|r|}
\hline
  \multicolumn{2}{|l||}{ \ } &
  \multicolumn{3}{c|}{Opt Gain (NYC)}\\
  \hline
$\alpha$ & \  Max \ & $k=2$  & $k=4$  & $k=8$   \\
  \hline
  \hline
2 & 7243 &  688.6  &  843.1  &  2088.8  \\
4  & 5133 & 798.9 & 1052.7 &  1270.9 \\
6 & 3080 & 209.6 & 506.6 & 762.5 \\
$\infty$ & 1449 & 40.1 & 77.1 & 313.4 \\
  \hline
\end{tabular}
\begin{tabular}{|r|r|r|}
  \hline
  \multicolumn{3}{|c|}{Approx Gain (NYC)}\\
  \hline
  $k=2$  & $k=4$  & $k=8$   \\
  \hline
  \hline
 688.6  &  789.2  &  1717.0   \\
 349.4 & 820.1 & 1224.3 \\
 209.6 & 458.3 &  734.6 \\
 40.1 & 75.7 & 217.2 \\
  \hline
\end{tabular}
\\
{ \ \ }
\\
\begin{tabular}{|l|r||r|r|r|r|r|r|}
\hline
  \multicolumn{2}{|l||}{ \ } &
  \multicolumn{3}{c|}{Opt Gain (FLA)}\\
  \hline
$\alpha$ & \  Max \ & $k=2$  & $k=4$  & $k=8$   \\
  \hline
  \hline
2 & 3448 &  1141.2 &  1287.6  &  1375.3  \\
4  & 2413 & 252.3 & 773.3 &  1001.4 \\
6 & 1426 & 128.8 & 318.0 & 457.7 \\
$\infty$ & 751 & 241.2 & 249.8 & 286.7 \\
  \hline
\end{tabular}
\begin{tabular}{|r|r|r|}
  \hline
  \multicolumn{3}{|c|}{Approx Gain (FLA)}\\
  \hline
  $k=2$  & $k=4$  & $k=8$   \\
  \hline
  \hline
 616.8  &  1178.0  &  1363.9   \\
 252.2 & 773.3 & 968.9 \\
 128.8 & 318.0 &  450.6 \\
 82.4 & 248.6 & 279.6 \\
  \hline
\end{tabular}

\end{scriptsize}
\end{center}
\vspace*{-4mm}
\caption{Gain values for MinMax on SF, NYC and FLA}
\vspace*{-4mm}
\label{tab:quality5}

\end{table}

\subsection{Effects of Parameters on Opt}

\begin{figure}[hbp]
  \centering
  \hspace*{-3mm}
    \includegraphics[width=0.51\columnwidth,height=1in]{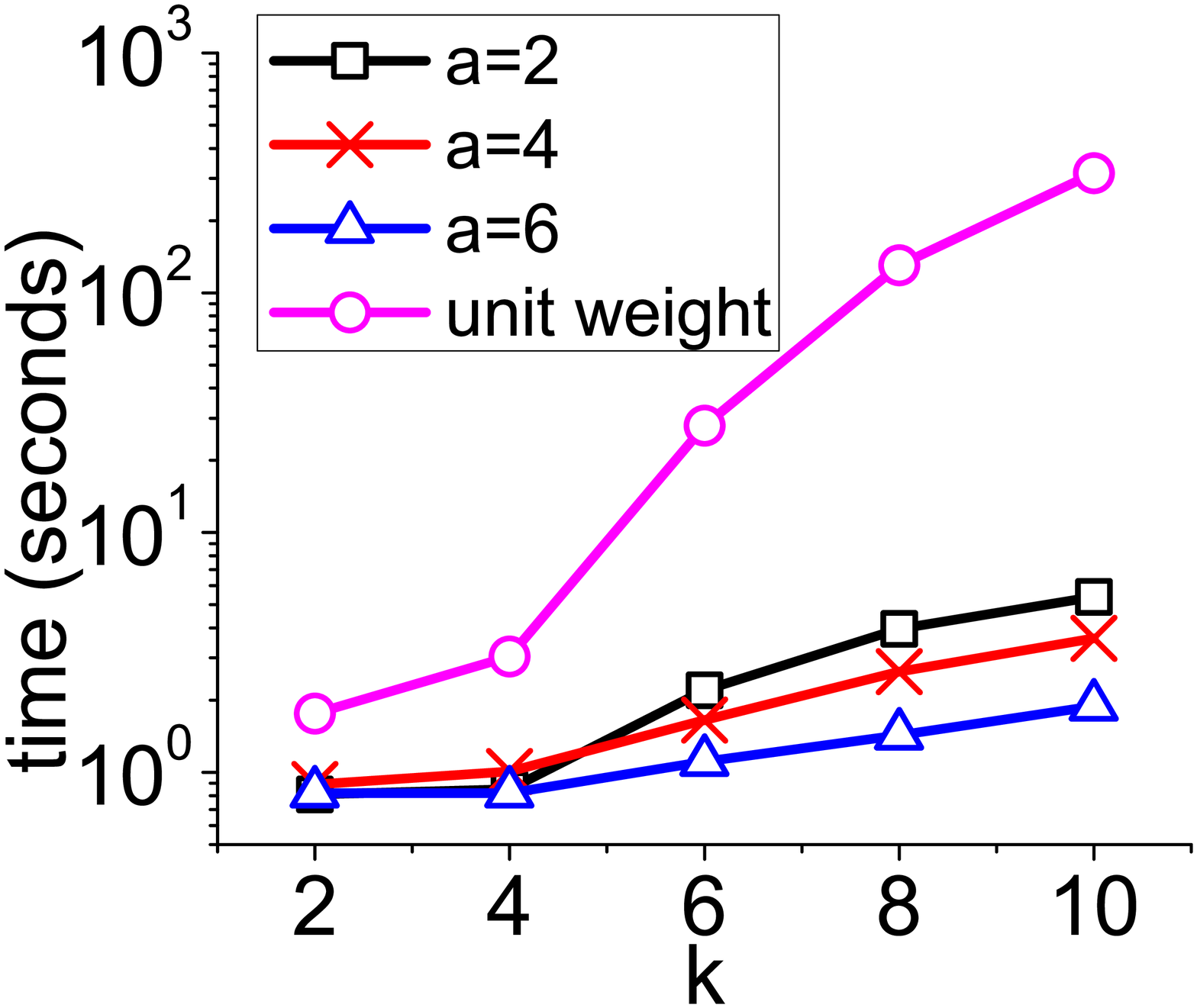}
    \hspace*{-2mm}
    \includegraphics[width=0.51\columnwidth,height=1in]{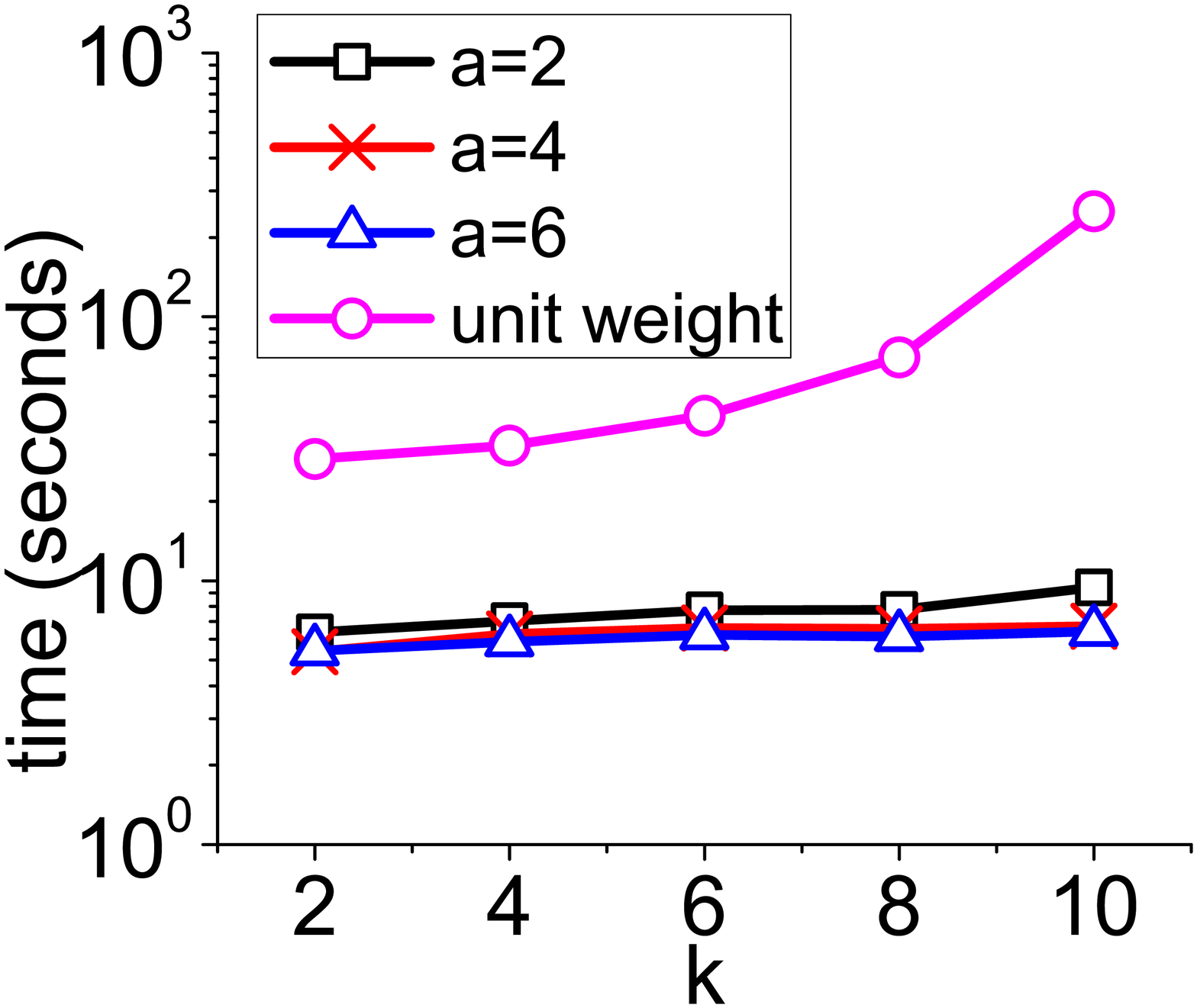}
    \\
    (a) \hspace*{1.5in} (b)
\caption{Effects of $k$ on the runtime for (a) SF and (b) FLA for MinMax
(a stands for $\alpha$ in the legend.)}
  \label{fig:minmax_k_Time}
  \vspace*{-3mm}
\end{figure}

We analyze the effect of $k$ on the runtime of the optimal algorithm in Figure \ref{fig:minmax_k_Time}. There are two different trends in the results.
If each client has a unit weight,
the runtime increases exponentially with $k$. When $\alpha \neq \infty $,
 meaning that the weights of the clients are not uniform, the runtime increases smoothly and it is easy to find the new server locations. This is because the new servers should be near to some clients with heavy weights.

\begin{figure}[h]
  \centering
  \hspace*{-3mm}
    \includegraphics[width=0.51\columnwidth,height=1in]{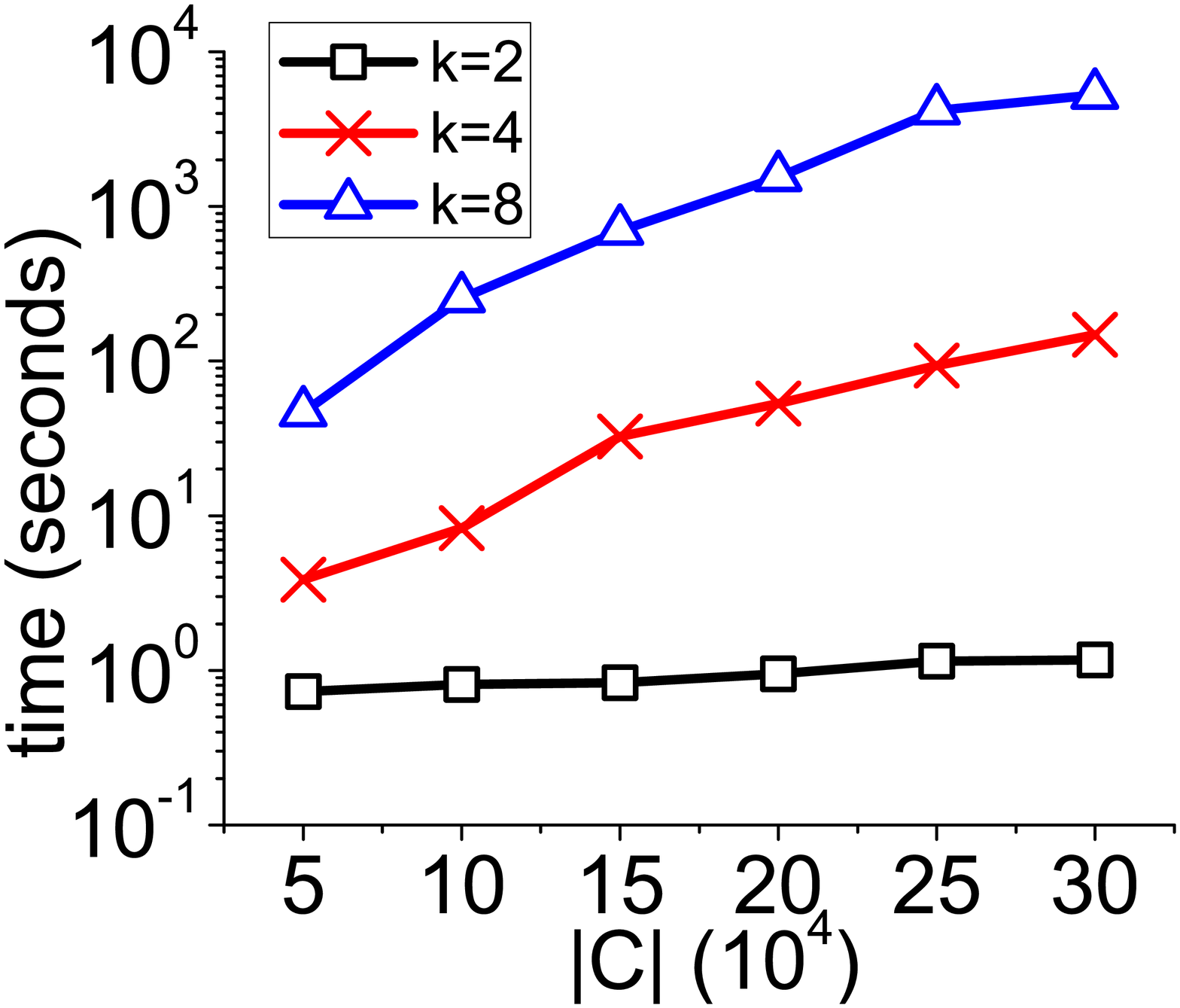}
    \hspace*{-2mm}
    \includegraphics[width=0.51\columnwidth,height=1in]{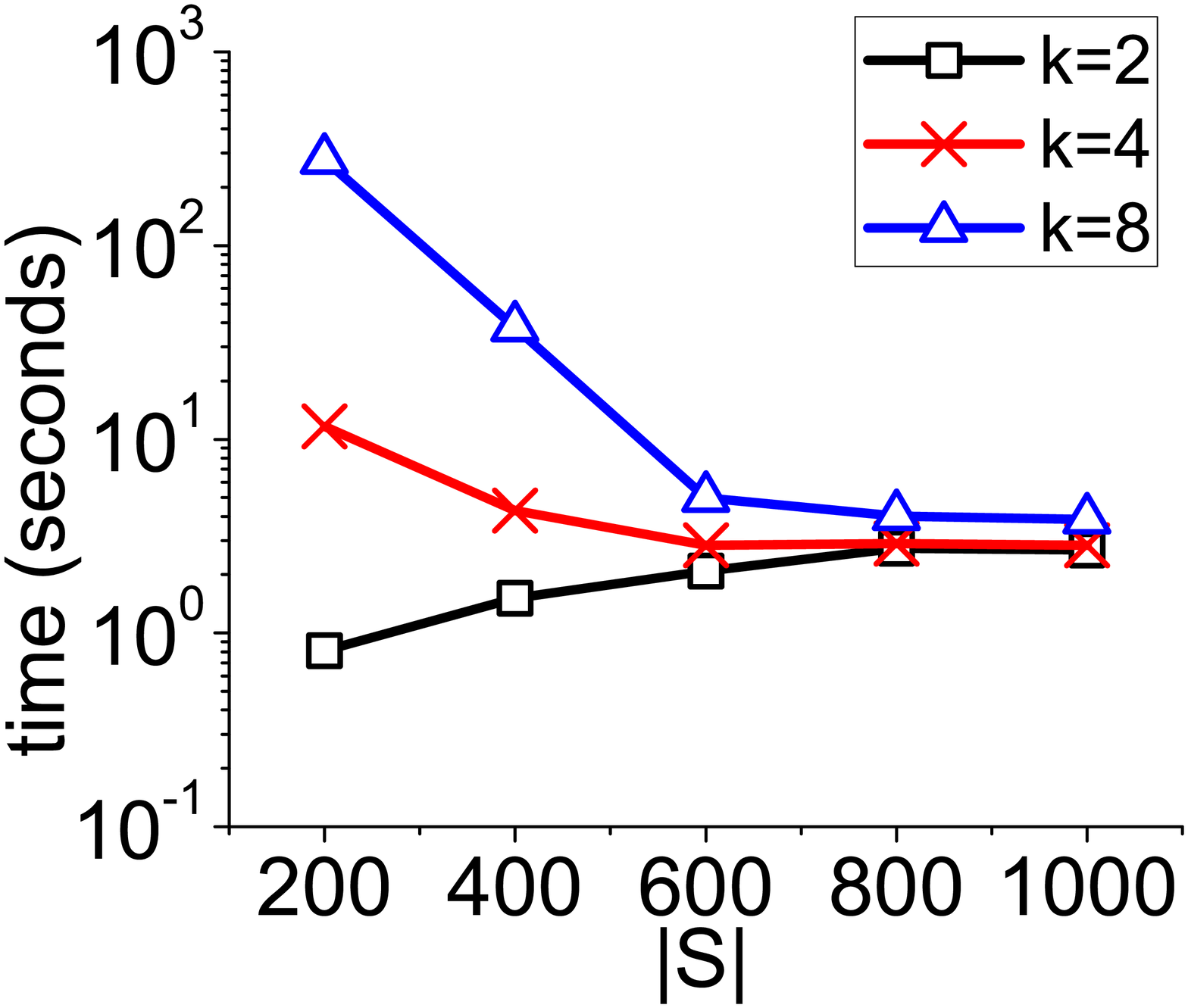}
    \\
    (a) \hspace*{1.5in} (b)
    \\
    \hspace*{-3mm}
      \includegraphics[width=0.48\columnwidth,height=1in]{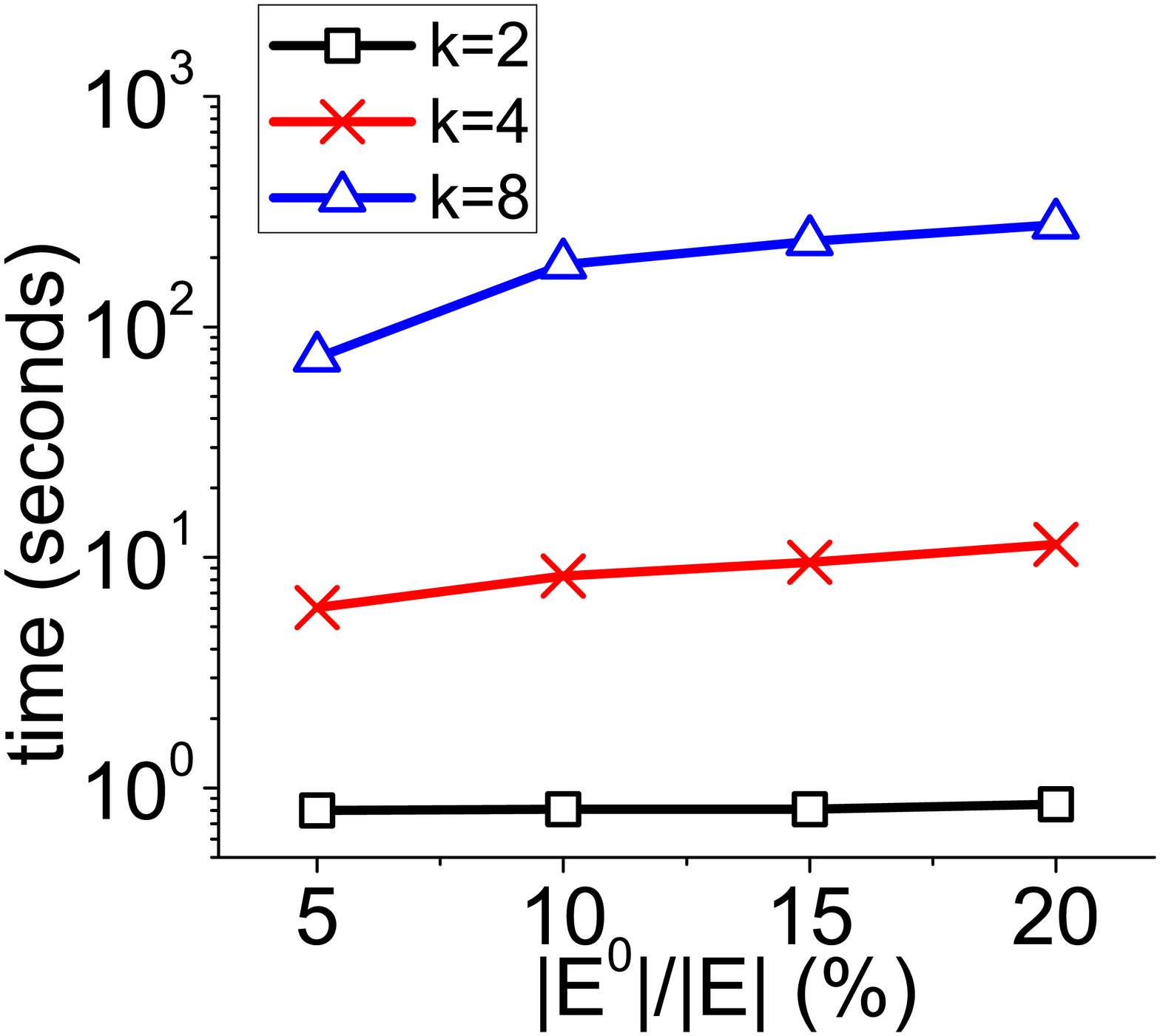}
      \hspace*{-2mm}
    \includegraphics[width=0.48\columnwidth,height=1in]{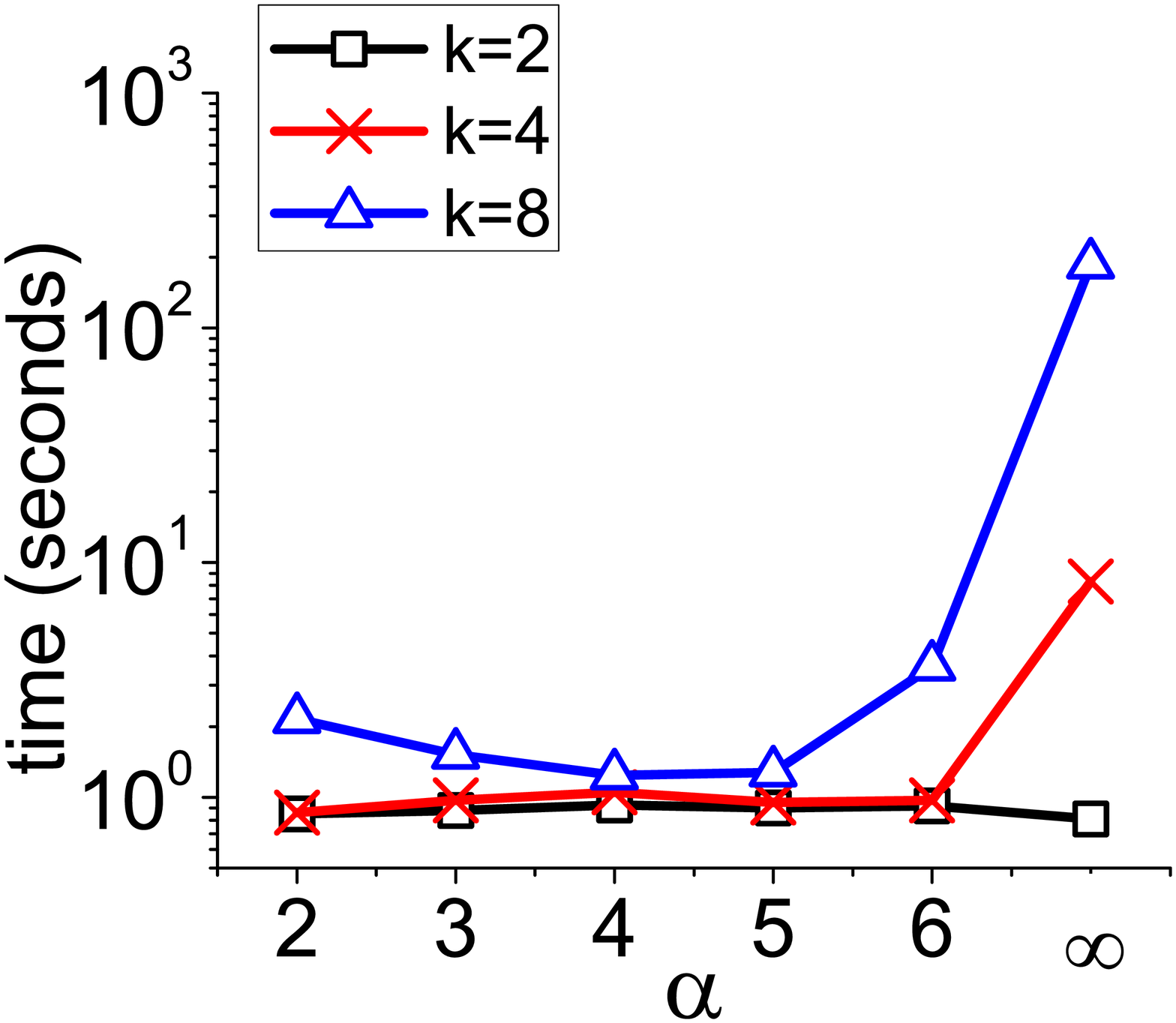}
    \\
    (c) \hspace*{1.5in} (d)
    \\
  \vspace*{-3mm}
\caption{Effects of $|C|$, $|S|$, $|E^0|$, and $\alpha$ for SF}
  \label{fig:minmaxTime}
  \vspace*{-3mm}
\end{figure}

We measure the effects of different parameters on the runtime of the optimal algorithm.
We study the effect of $|C|$ in Figure \ref{fig:minmaxTime}(a). The runtime increases with $|C|$. In the experiment, the number of servers and their positions are fixed. The optimal algorithm result and $Gain$ do not change much as $|C|$ increases. The number of clients processed increases when $|C|$ increases. 

The effect of $|S|$ is shown in Figure \ref{fig:minmaxTime}(b). The runtime increases with $|S|$ when $k$ is small and decreases when $k$ is large. The runtime depends on two factors: the time for building $NLC$s($CCL$s) and computing $PSP$s, and the time for processing $PSP$s. The first factor dominates when $k$ is small. When $|S|$ is small, the cost difference between two consecutive clients in the sorted list is large, so we can find the result quickly and the second factor is not significant. 
When $k$ is large, the second factor dominates. When $|S|$ is small, the $NLC$s of clients are large and there are more $PSP$s. Since the sizes of $NLC$s and number of $PSP$s decrease as $|S|$ increases, the runtime decreases when $k$ increases.


In Figure \ref{fig:minmaxTime}(c), we study the effect of $|E^0|$.
We vary $|E^0|/|E|$. 
The time increases steadily with the increase in $|E^0|$.
The pruning of $NLC$ computation and $PSP$s is highly effective. In all our experiments, the number of $NLC$s that are computed is a few hundred
at most, and the number of $PSP$s that are not pruned is not more than twenty. The increase in runtime is gentle due to the effective pruning.
%
%
In Figure \ref{fig:minmaxTime}(d), we study the effect of Zipf parameter $\alpha$. Skewness in weights is beneficial to the runtime.
Intuitively, when there exist some heavy weight clients, it becomes easier to select the new server locations, since they should be near to such clients. This sharpens the pruning effects, and the running time of Opt is similar to that of Approx when $\alpha \neq \infty $ (see Table \ref{tab:quality4}).

\begin{figure}[h]
  \centering
  \hspace*{-3mm}
    \includegraphics[width=0.51\columnwidth,height=1in]{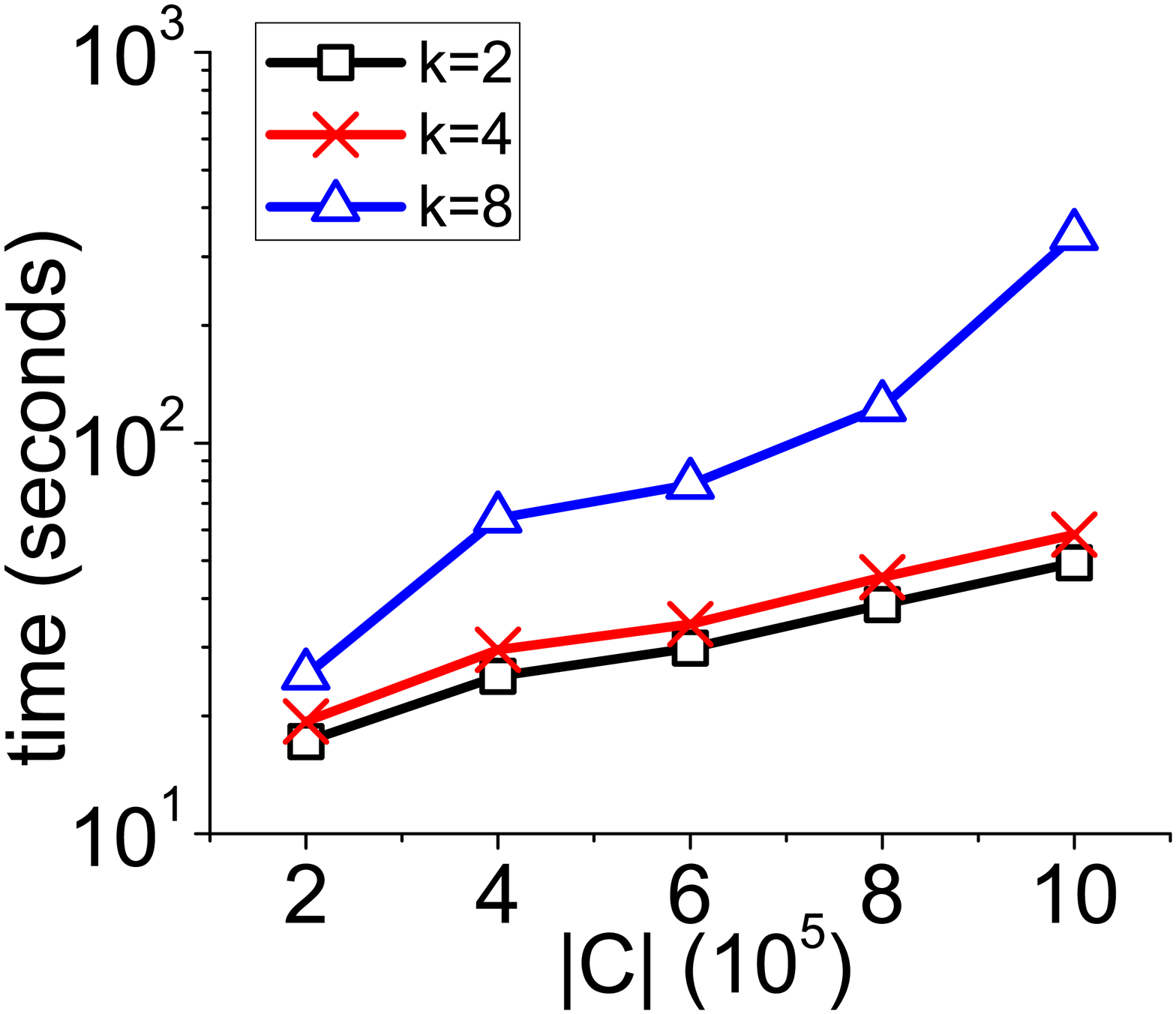}
    \hspace*{-2mm}
    \includegraphics[width=0.51\columnwidth,height=1in]{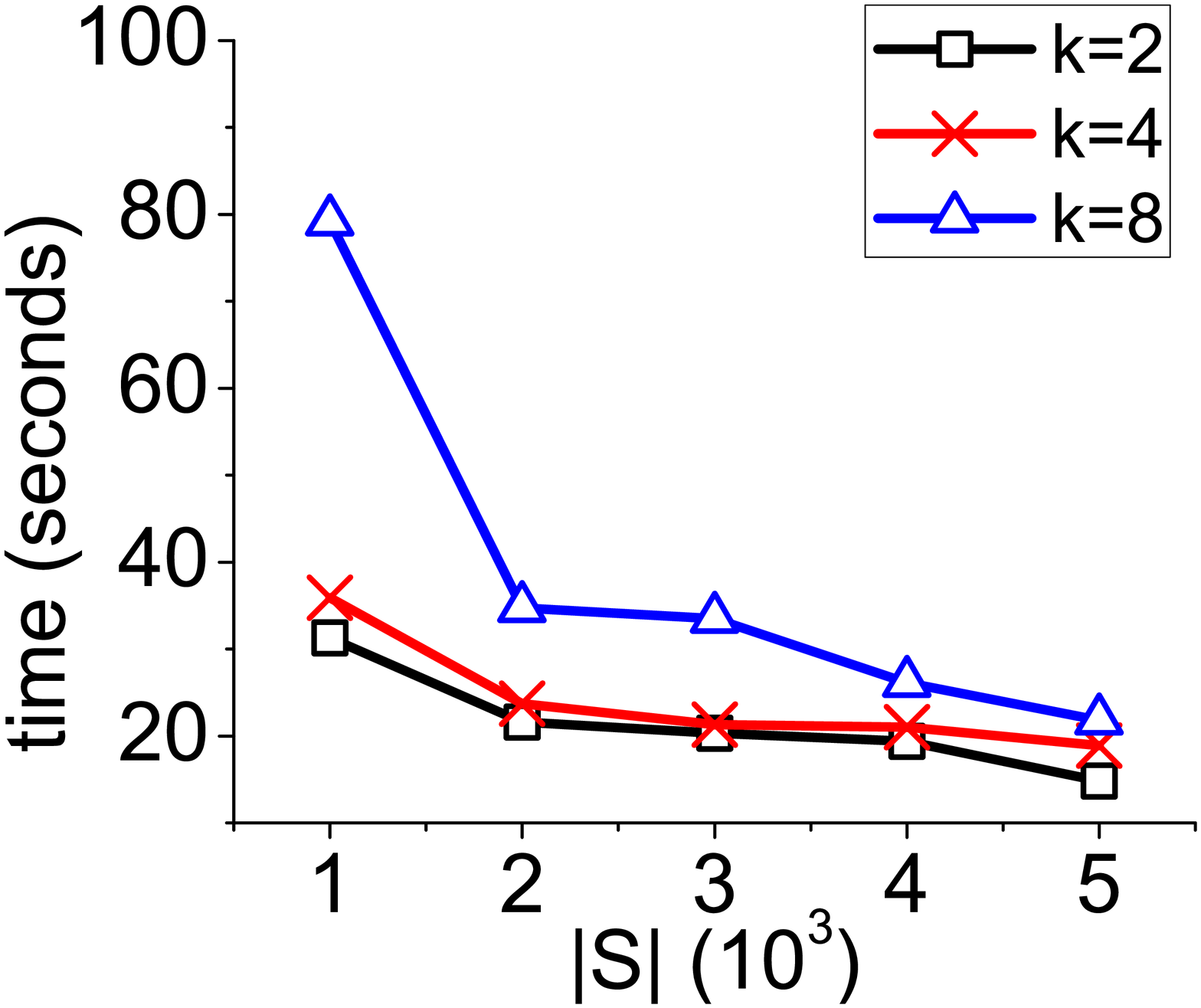}
    \\
    (a) \hspace*{1.5in} (b)
    \\
    \hspace*{-3mm}
      \includegraphics[width=0.51\columnwidth,height=1in]{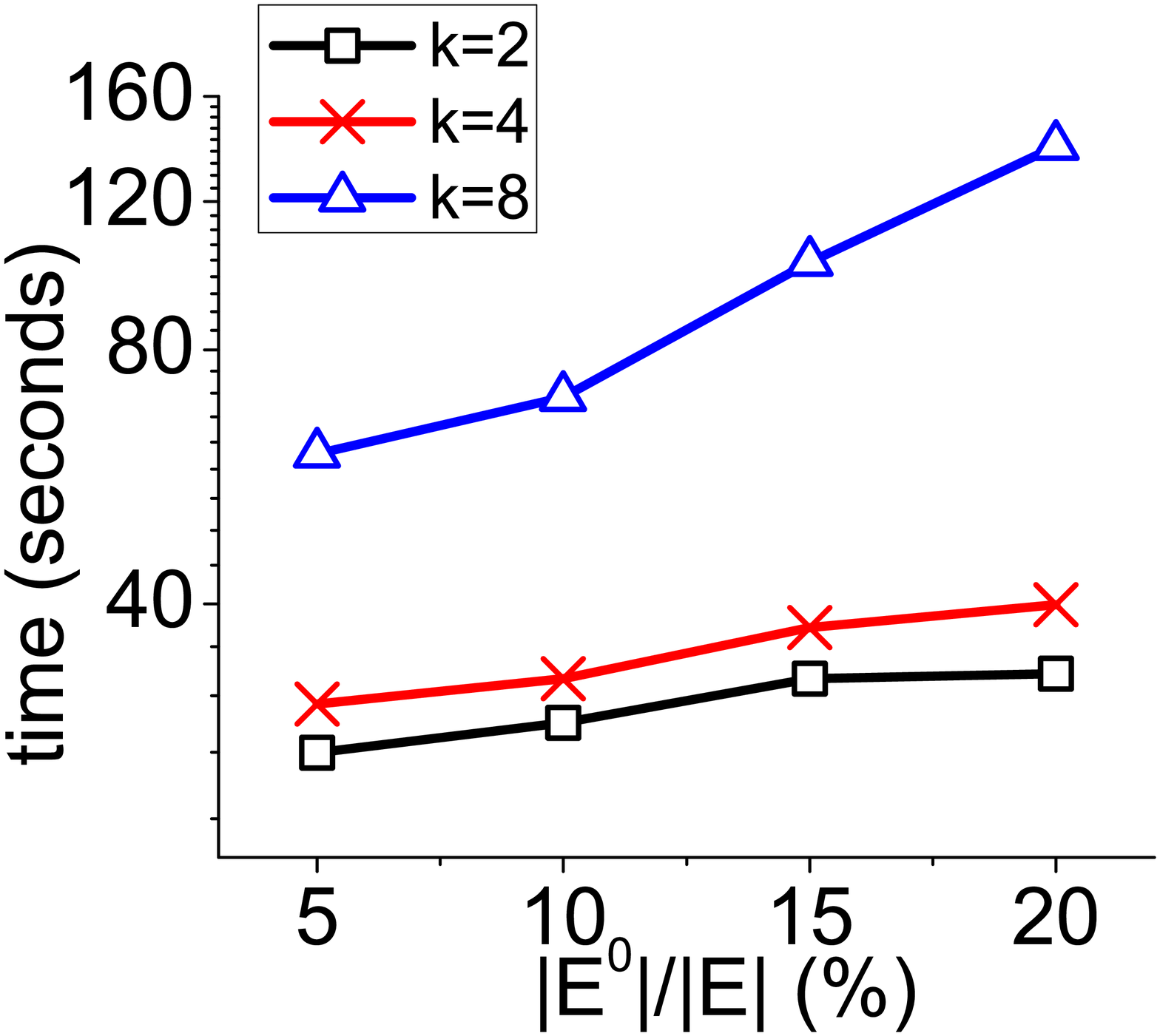}
      \hspace*{-2mm}
    \includegraphics[width=0.51\columnwidth,height=1in]{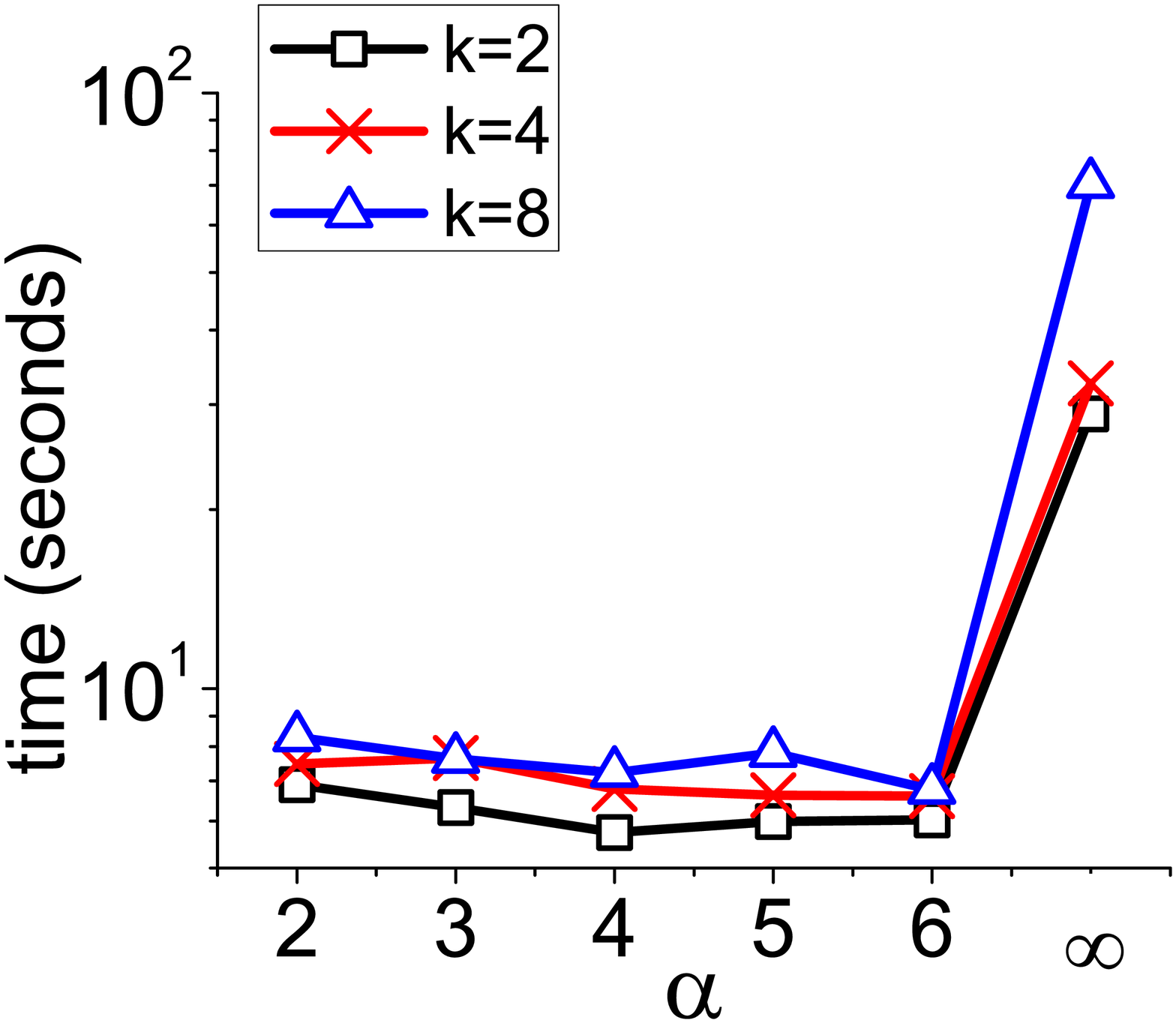}
    \\
    (c) \hspace*{1.5in} (d)
    \\
  \vspace*{-3mm}
\caption{Effects of $|C|$, $|S|$, $|E^0|$, and $\alpha$ for FLA}
  \label{fig:minmaxTime2}
  \vspace*{-1mm}
\end{figure}


A similar set of results has been obtained for NYC. For FLA, the results are shown in Figure \ref{fig:minmaxTime2}.
The trends are mostly similar to SF, except that in Figure \ref{fig:minmaxTime2}(b), the time decreases with $|S|$ for all values of $k$. This is because FLA is a state where locations are more spread out compared to a city. Even with $k=2$, many clients are covered by the new servers, so the time for handling the $PSP$s and $k$-candidates dominates, and this decreases as the $NLC$ sizes decrease with increasing $|S|$.

In summary, $Opt$ can handle problems with small $k$ values efficiently. $Approx$ takes less time than $Greedy$ and returns a solution with a much better quality. Thus, we improve on the status quo where $k=1$ is solved with an optimal solution and the cases of $k > 1$ are solved by $Greedy$.




\subsection{Effects of the Enhancement Strategies}

We measure the effects of the enhancement strategies of optimal algorithm QuickMinMax in Section \ref{sec:enhancedMinMax}.
We show the number of iterations terminated early by $Strategy$ $1$ and $Strategy$ $2$ in Section \ref{sec:strategy} in Table \ref{tab:strategy}. On average, we terminate early and jump to the next iteration with Strategies 1 and 2 in over 99.25\% of the iterations.

\begin{table}[h]
\begin{center}
\begin{scriptsize}
\begin{tabular}{|l||r|r|r|r|r|}
  \hline
k (SF) & 2 & 4 & 6 & 8 & 10   \\
  \hline
  \hline
Total iterations &	43	&	93	&	528	&	700	&	777 \\
Strategy 1  &\hspace{0.20cm} 5 &\hspace{0.20cm} 25 & \hspace{0.20cm} 373 & \hspace{0.20cm} 478 & \hspace{0.20cm} 512  \\
Strategy 2  & 37 &  67 & 153 & 218 & 263  \\
Strategies 1+2(\%)    & 97.68 &  98.92 & 99.62 & 99.43 & 99.74  \\
\hline
\hline
k (FLA) & 2 & 4 & 6 & 8 & 10   \\
  \hline
  \hline
Total iterations  & 242 &  278 & 349 & 409 & 444  \\
Strategy 1  & 179 & 214 & 282 & 331 & 347  \\
Strategy 2  & 62 &  63 & 66 & 76 & 91  \\
Strategies 1+2(\%) & 99.59 &  99.64 & 99.71 & 99.51 & 98.65  \\
\hline
\end{tabular}
\end{scriptsize}
\vspace*{-2mm}
\caption{Number of iterations terminated early}
\vspace*{-3mm}
\label{tab:strategy}
\end{center}
\end{table}

\begin{table}[h!]
\begin{center}
\begin{scriptsize}
\begin{tabular}{|l||r|r|r|r|r|}
  \hline
k (SF) & 2 & 4 & 6 & 8 & 10   \\
  \hline
  \hline
Total 	    & 1247 &  3712 & 58458 & 78530 & 81645 \\
Strategy 4  & \hspace{0.20cm} 37 &  \hspace{0.20cm} 108 & \hspace{0.20cm} 543 & \hspace{0.20cm} 1512 & \hspace{0.20cm} 1597  \\
Strategies 4+3  & 3 & 5 & 8 & 17 & 20 \\
PSPs pruned (\%) & 99.75 &  99.86 & 99.98 & 99.98 & 99.98  \\

\hline
\hline
k (FLA) & 2 & 4 & 6 & 8 & 10   \\
  \hline
  \hline
Total 	     & 46580 &  77079 & 144833 & 229004 & 229414  \\
Strategy 4   & 228 &  488 & 644 & 1710 & 2660  \\
Strategies 4+3  & 3 & 5 & 7 & 9 & 10 \\
PSPs pruned (\%) & 99.99 &  99.99 & 99.99 & 99.99 & 99.99  \\
 \hline
\end{tabular}
\\
\end{scriptsize}
\vspace*{-2mm}
\caption{Number of PSPs after pruning by $Strategies$ $3$ and $4$} 
\vspace*{-3mm}
\label{tab:pruning}
\end{center}
\end{table}

\begin{table}[h!]
\begin{center}
\begin{scriptsize}
\begin{tabular}{|l||r|r|r|r|r|}
  \hline
k (SF) & 2 & 4 & 6 & 8 & 10   \\
  \hline
  \hline
SF & 2 &  4 & 27 & 12316  & 141008 \\

\hline
\hline
FLA & 2 &  4 & 6 & 9 & 44 \\
 \hline
\end{tabular}
\\
\end{scriptsize}
\vspace*{-2mm}
\caption{Number of $k$-candidates computed}
\vspace*{-2mm}
\label{tab:kcandidate}
\end{center}
\end{table}


Table \ref{tab:pruning} shows the total number of PSPs computed in the road network, number of PSPs left after $Strategy$ $4$ and number of PSPs left after both $Strategies$ 4 and $3$ in the last iteration. With $Strategy$ $4$ and $Strategy$ $3$, we reduce the $PSP$ number by 99.90\% on average. Only very few PSPs are used for computing $k$-$candidate$s. Table \ref{tab:kcandidate} shows the number of $k$-$candidate$s computed in the optimal algorithm. Given a set $X$ of $PSP$s in each iteration, we only compute the new and updated $k$-$candidate$s. In addition, $PSP$s are pruned by different strategies, thus the number of $k$-$candidate$s computed is much smaller than the bound of $O(|X|^k)$.
%
%
These results show that our enhancement strategies lead to significant improvements in the overall performance.

\section{Conclusion}
\label{sec:concl}

We consider the problem of MinMax for finding multiple optimal minmax locations on a road network. We propose a novel algorithm based on the concepts of client cost lines and potential server points, with search space pruning and early stopping strategies.
Our empirical study shows that our proposed algorithm generates significantly better solutions
compared to previous works on three real road networks.
Other OLQ problems with multiple new servers will be interesting problems for further studies. 

\bibliographystyle{abbrv}

{\small


\bibliography{ref}
}

\begin{appendix}




{\small PROOF OF LEMMA 3}:
$\texttt{zmax}(C',p) = \max(p.cost, Cost_S(C''))$,
where $C''$ is the set of clients in $C'$ not covered by $p$, and
$Cost_S(C'') = \max( Cost_S(c)|c \in C'')$.
When $C'$ is the set of clients covered by $p$, $C'' = \emptyset$.
Hence,
$\texttt{zmax}(C',p)=p.cost$.
Also, since $\texttt{zmax}(C',p) \geq \texttt{zmax}(C',q) \geq q.cost$, thus, $q.cost \leq p.cost$.
Since $p \in A$, and $\texttt{amax}(A)$ is lower bounded by the costs of the apex points in $A$,
$\texttt{amax}(A) \geq p.cost$.
For a client $c'$ covered by $q$, we have
$Cost_{S \cup A'}(c') \leq q.cost \leq  p.cost$ $\leq \texttt{amax}(A)$.
For a client $c'$ not covered by $q$, it is either covered
by $p$ or not covered by $p$.
If $c'$ is not covered by $p$,
since $A' - \{q\} = A - \{p\}$, then
$Cost_{S \cup A'}(c') = Cost_{S \cup A}(c')$.
If $c'$ is covered by $p$, then
$c' \in C'$.
Given that $\texttt{zmax}(C',p) \geq \texttt{zmax}(C',q)$,
$\texttt{zmax}(C',q) \leq p.cost$.
Since $c'$ is not covered by $q$,
it must hold that $Cost_S(c') > q.cost$. We deduce that
$\texttt{zmax}(C',q)$
$\geq$
$\max(q.cost, Cost_S(c'))$ = $Cost_S(c')$.
Thus, $Cost_S(c') \leq \texttt{zmax}(C',q) \leq p.cost$.
If $c'$ is covered by some point in $S \cup (A \cap A')$, then
$Cost_{S \cup A'}(c') = Cost_{S \cup A}(c')$,
otherwise, $c'$ is not covered by any point in $S \cup A'$,
and $c'$ is covered only by $p$ in $S \cup A$.
We have $Cost_{S \cup A'}(c') = $
$Cost_{S \cup A}(c') = Cost_S(c')$.
Note that $\texttt{amax}(A) \geq Cost_{S \cup A}(c)$ for all $c \in C$.
Thus, considering all cases for any given client $c'$,
$Cost_{S \cup A'}(c') \leq \texttt{amax}(A)$.
We conclude that $\texttt{amax}(A') \leq \texttt{amax}(A)$
and the lemma holds.
\done

\vspace*{10pt}



\begin{figure}[htbp]
\begin{center}
\hspace*{-2mm}
\includegraphics[width=3.3in]{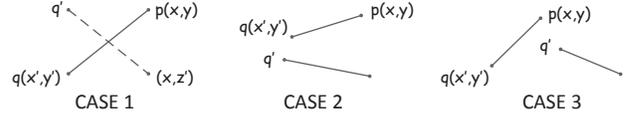}
\vspace*{-5mm}
\end{center}
\caption{3 cases in the proof of Lemma \ref{lem:PSP1}}
\vspace{-2mm}
\label{fig:lem1}
\end{figure}

\noindent
{\small PROOF OF LEMMA 4}:
Let $\ell = [a,b]$. Assume $I(\ell) \neq \emptyset$.
If a point $p = (x,y)$ on $CCL_\ell(c)$ is not in $I(\ell)$, and $p \not\in B(\ell)$, then there must be a point $q=(x',y')$ with $q.cost < p.cost$
(i.e., $y' < y$)
on $CCL_\ell(c)$ closest to $p$ which is either a lowest boundary point
or $q \in I(\ell)$.
We show that replacing $p$ by $q$ does not increase the overall minmax cost.
Clearly, we only need to consider clients covered by $p$.
There are different cases for such a covered client $c'$.
We show in each case that $\texttt{zmax}(\{c'\},p) \geq$ $\texttt{zmax}(\{c'\},q)$.

\noindent
[CASE 1]:
Suppose $p$ covers a client $c'$ where $CCL_\ell(c')$ contains no end point.
We show that $q$ also covers $c'$.
If this is true, then
since $q.cost < p.cost$,
$\texttt{zmax}(\{c'\},p) \geq$ $\texttt{zmax}(\{c'\},q)$.
%
Assume on the contrary that $p$ covers such a client $c'$ and $q$ does not.
Recall that $p = (x,y)$, $q = (x',y')$.
Without loss of generality,
let $x' < x$.
Since $q$ cannot cover $c'$, there exists a point
$q'=(x', z)$ on $CCL_\ell(c')$ with $z > y'$.
This is illustrated in Figure \ref{fig:lem1} as CASE 1.
Since $p$ covers $c'$,
we have
$p' = (x,z') \in CCL_\ell(c')$, where $z' < y$.
However, given the points $p,q$, $p',q'$, there must exist an intersecting point $(\hat{x},\hat{y})$ of $CCL_\ell(c)$ and $CCL_\ell(c')$, such that either $x'<\hat{x}<x$, or
$x < \hat{x} < x'$,
which contradicts the assumption that $q$ is the closest point to $p$ in $I$. Thus, $q$ also covers $c'$.

\noindent
[CASE 2]:
Next, suppose an end point $q'$ of $CCL_\ell(c')$ is in $(a,b)$ in $\ell$ and $Cost_S(c') \leq y'$. I.e. $q'$ is not a boundary of $\ell$. Since $y' < y$, clearly,
$\texttt{zmax}(\{c'\},p) \geq$ $\texttt{zmax}(\{c'\},q)$.
See Figure \ref{fig:lem1} CASE 2 for a possible scenario.

\noindent
[CASE 3]:
Finally, consider the case where an end point $q'$ of $CCL_\ell(c')$ is in $(a,b)$ in $\ell$ and $Cost_S(c') > y'$.
Since $p$ covers $c'$,
$Cost_S(c') \leq y$.
Thus,
$\texttt{zmax}(\{c'\},q)$ = $Cost_S(c') \leq$
$\texttt{zmax}(\{c'\},p)$. See Figure \ref{fig:lem1} CASE 3 for an illustration.

From all the above cases, we conclude that for the set $C'$ of clients covered by $p$,
$\texttt{zmax}(C',p) \geq \texttt{zmax}(C',q)$.
The claim thus follows from Lemma \ref{lem:minmax}.
\done 
\end{appendix}

\end{document}